\documentclass[preprint,12pt, 5p,times,twocolumn]{elsarticle}
\usepackage[T1]{fontenc}
\usepackage{graphicx}
\usepackage{amsmath}
\usepackage{mathtools}
\usepackage{multirow}
\usepackage{color, soul}
\usepackage{hyperref}

\journal{Kowledge-Based Systems}

\begin{document}
	
\begin{frontmatter}
	
\title{A Predictive Neural Network Architecture for Early Detection of Low-Rate Cyberattacks
\tnoteref{label1}
}
\tnotetext[label1]{This preprint is published at Knowledge-Based Systems, DOI: \href{https://doi.org/10.1016/j.knosys.2026.115995}{10.1016/j.knosys.2026.115995} .}

\author[IITIS]{Mert Nak{\i}p\corref{cor1}}
\ead{mnakip@iitis.pl}
\cortext[cor1]{Corresponding author.}
\affiliation[IITIS]{organization={Institute of Theoretical and Applied Informatics, Polish Academy of Sciences},
	addressline={Baltycka 5}, 
	city={Gliwice},
	postcode={44--100}, 
	country={Poland}}

\begin{abstract}
Low-Rate Denial of Service (LDoS) attacks pose a significant challenge to IoT networks due to their subtle and prolonged nature, often evading traditional intrusion detection systems. This paper presents IDQS (Intrusion Detection via QoS Prediction), a lightweight and proactive framework for early LDoS attack detection. IDQS integrates two new key components: (i) RTP-QoS, a Recurrent Trend Predictive Neural Network that learns and forecasts future Quality of Service (QoS) based on historical traffic patterns, and (ii) PDM, a Pairwise Decision Model that evaluates discrepancies between predicted and actual QoS to identify potential attacks. Evaluated on the public SDN-SlowRate-DDoS and CIC-IDS2017 datasets, IDQS respectively achieves over 79\% and 91\% detection accuracy across most attack scenarios with high recall and low false negatives, while maintaining an end-to-end inference time of just 0.28 seconds. The results demonstrate the effectiveness and efficiency of IDQS for real-time deployment in resource-constrained IoT environments.
\end{abstract}

\begin{keyword}
Quality of Service, Prediction, Trend Predictive Neural Network (rTPNN), Internet of Things (IoT), Low-Rate Denial of Service (LDoS)
\end{keyword}

\end{frontmatter}

\section{Introduction}

The Internet of Things (IoT) is rapidly evolving with an increasing number of device uses and application areas. IoT devices and the networks they create --especially when integrated with Artificial Intelligence (AI) algorithms-- are frequently used to automate systems in various application areas \cite{al2019artificial, baccour2022pervasive, manoharan2021artificial}. However, these devices and networks, often deployed at large scale \cite{EricssonMobilityReport2019}, not only contain numerous vulnerabilities \cite{HP} but are also easily affected by collision and congestion \cite{gelenbe2022improving, rodoplu2020multiscale}. Therefore, they become a direct target of cyberattacks and a gateway for cybercriminals to reach critical infrastructures.

Denial of Service (DoS) and Distributed Denial of Service (DDoS) attacks are two of the most common cyberattacks posing a significant threat to IoT. Indeed, in the second quarter of 2022, DDoS attacks annually occurred 109\% more in the network layer and 72\% more in the application layer \cite{ddos_stats_2022}. These attacks are often straightforward and implemented with the aim of disrupting the normal services of the targeted IoT system. To this end, they usually send a large number of requests in a fairly short period of time, so much so that the target system cannot cope with these requests and is overwhelmed \cite{dos_definition}. Although successful DoS or DDoS attacks have devastating impacts and serious consequences, state-of-the-art security measures \cite{nakip2021mirai, LCN, Google_DDoS_2023} can detect and mitigate such attacks successfully (but still raising various performance-related issues). 

On the other hand, cyberthreats are also evolving and becoming more complex along with cybersecurity methods. For the security of IoT networks, which are often prone to network performance loss, e.g. Quality of Service (QoS) degradation, Low-Rate DoS (LDoS) attacks, characterized by their subtle and sophisticated nature, pose a significant challenge to traditional detection methods. These attacks often involve the exploitation of application vulnerabilities or the manipulation of network traffic at a level that evades conventional detection mechanisms \cite{Zhijun2020Survey}. Due to their low resource requirements with low and slow characteristics, these attacks can easily be launched and remain undetected for a long time \cite{cloudflare}. Since traces of LDoS attacks cannot be sharply distinguished from normal network traffic, detecting and preventing these attacks is very challenging. {In practice, existing detection methods predominantly rely on static classification or fixed thresholding, which fail to account for the inherent stochasticity of IoT network traffic. This creates a critical gap, as such systems are often unable to differentiate between a deliberate LDoS attack and legitimate network fluctuations or transient congestion. Consequently, there is a practical need for an adaptive mechanism that can model the expected dynamic state of the network to isolate anomalous signatures in real-time.} Therefore, the development of robust and adaptive detection techniques tailored specifically for low-rate attacks is essential.

{This paper presents a system-level holistic design named Intrusion Detection via QoS Prediction (IDQS) for identifying the almost invisible footprint of LDoS attacks. While traditional systems rely on volume-based spikes, IDQS is a multi-stage predictive pipeline designed to detect the subtle degradation that stays under the radar of absolute-threshold models. The proposed IDQS is comprised of two main components: Recurrent Trend Predictive QoS Predictor (RTP-QoS) to accurately forecast the network QoS and Pairwise Decision Model (PDM) to effectively identify malicious traffic. By shifting the detection approach from a simple classification task to a predictive discrepancy analysis, the framework accounts for slow temporal drifts in QoS.} Accordingly, the key contributions of this paper are as follows: 

\begin{itemize} 
	\item \textbf{{System-Level Multi-Stage Pipeline}}: {The proposed IDQS fills the gap between existing static classification models and dynamic network behavior via a three-stage pipeline: (i) extraction of relevant traffic features, (ii) forecasting future QoS via RTP-QoS, and (iii) identifying attack patterns via a predictive discrepancy analysis in the PDM.}
	\item \textbf{Specialized QoS-Predictive Architecture}: {We develop the RTP-QoS model, which introduces a new QoS prediction layer for the rTPNN architecture. While the rTPNN layer is used for trend extraction, this specific synchronization enables the system to learn and anticipate future QoS patterns based on expected normal behavior. The internal structure of RTP-QoS predicts upcoming traffic measurements and trends, while the output layer incorporates a recurrent error correction feature to predict QoS.} 
	\item {\textbf{Early and Accurate Detection via PDM as a Comparator Layer}: The PDM developed in this paper acts as a specialized comparator layer. It is designed to first capture the relationships between the predicted and actual QoS pairs and to process them together with the forecast QoS to identify signatures of low-rate attacks that evade traditional detection.} 
	\item {\textbf{Practical Efficiency for IoT Gateways}: We demonstrate that IDQS is a significant system-level contribution for resource-constrained environments. With an average inference time of 0.28 seconds, the framework provides a proactive defense mechanism capable of running on low-power IoT gateways in real-time.} 
\end{itemize}

The performance of the proposed IDQS is evaluated using public ``SDN-SlowRate-DDoS'' \cite{dataset} and ``CIC-IDS2017'' \cite{CCIDS2017_dataset}. The results reveal that IDQS achieves an accuracy greater than 0.79 in the majority of experiments conducted during 5-fold cross-validation. In addition, the RTP-QoS and PDM components of IDQS are compared against the well-known neural networks, namely Long Short-Term Memory (LSTM), Multilayer Perceptron (MLP), Nonlinear Perceptron (NP), and the Feed Forward Network. The results demonstrate that IDQS significantly outperforms the neural networks compared as it achieves superior prediction accuracy (N-SMAPE < 1\%) and more reliable detection accuracy. {Furthermore, the computational efficiency of IDQS ensures it can be effectively deployed in real-world IoT infrastructures without incurring significant overhead.}




The remainder of this paper is organized as follows:  
Section~\ref{sec:Method} details IDQS respectively presenting the traffic metrics, the RTP-QoS model, and the PDM for detection of LDoS attack. Section~\ref{sec:Results} evaluates the performance of the IDQS on a public dataset and discusses the experimental results. Section~\ref{sec:Conclusion} summarize the paper and provides insights into the future implications of the proposed methodology.

\section{{Related Works}}\label{sec:RelatedWorks}


\subsection{{High-Rate Intrusion Detection}}

Earlier research and recent works are mostly focused on detecting attacks with high traffic rate and intense impact on the system resources in a short period of time, such as Botnet attacks. For instance, in Reference~\cite{bovenzi2023network}, Bovenzi et al. applied Deep Learning (DL) models --specifically Autoencoders (AEs) and KitNET-- for unsupervised early anomaly detection within IoT datasets, namely IoT-23~\cite{iot23dataset} and Kitsune~\cite{kitsune_dataset}. The findings presented in~\cite{bovenzi2023network} highlighted the feasibility of early detection of IoT network attacks, demonstrating that the analysis of just the first four packets yielded the most effective detection performance. References \cite{MIRAI} and \cite{Incremental} introduced a lightweight Intrusion Detection System (IDS) for detecting Mirai Botnet attacks, which analyzes network traffic using three metrics designed to quantify traffic density from multiple perspectives. The IDS was trained in an unsupervised manner, both offline~\cite{MIRAI} and in a quasi-online incremental fashion~\cite{Incremental}, utilizing only normal traffic to construct an auto-associative memory. In order to detect Botnet attacks in the N-BaIoT dataset, Reference~\cite{Htwe} employed Classification and Regression Trees (CART) in conjunction with feature selection, while Sriram et al.\cite{Sriram} conducted a comparative analysis involving seven distinct Machine Learning (ML) algorithms, including Naive Bayes (NB), K-Nearest Neighbors (KNN), and Support Vector Machines (SVM). In addition, Reference~\cite{Soe} proposed a Botnet detection framework consisting of two sequential phases: the first phase focuses on training the selected ML model and performing feature selection, and the second phase executes the actual attack detection. Within this architecture, the authors utilized both MLP and NB, assessing their performance on the N-BaIoT dataset. In recent work \cite{nakip2024online}, in order to enable fully online learning of IDS parameters, a novel Self-Supervised Intrusion Detection (SSID) framework was proposed. SSID continuously learns from incoming network traffic packets and evaluates the trustworthiness of the IDS by assessing both its generalization capability and its accuracy on the traffic data used for learning. 

{In Reference~}\cite{polat2022novel}{, the vulnerabilities of SDN-based SCADA systems were addressed by introducing a Recurrent Neural Network (RNN) model utilizing parallel Long Short-Term Memory (LSTM) and Gated Recurrent Unit (GRU) architectures. Their approach, enhanced by transfer learning and SVM for feature classification, demonstrating the efficacy of deep recurrent structures in protecting critical industrial infrastructure from service disruptions. Building upon these architectural advancements, Reference~}\cite{polat2024hybrid}{ proposed a hybrid AI-powered detection module for Software Defined-based Vehicular Ad Hoc Networks (SD-VANETs), combining 1D Convolutional Neural Networks (1D-CNN) and Decision Trees to identify DDoS traffic with high precision.}

Furthermore, Reference~\cite{jain2025intelligent} developed an Intelligent Zero-day Attack Detection System using Unsupervised Machine Learning (ZdAD-UML) to enhance IoT network security by detecting previously unknown intrusions without requiring labeled data. The system combines Hippopotamus Optimization Algorithm for feature selection, Enhanced K-means Clustering-based Weighted Crayfish Auto Encoder for feature extraction, and Pufferfish Optimization-based threshold for anomaly detection.

{In contrast to the works which prioritize high-volumetric traffic signatures and rapid resource exhaustion, this paper focuses on detecting low-rate cyberattacks that subtly degrade system performance. While high-rate detection frameworks are optimized for binary connectivity alarms and massive packet flows, the proposed IDQS framework addresses the stealthier challenge of identifying anomalies that remain hidden within seemingly normal traffic volumes.}

\subsection{{Low-Rate Intrusion Detection}}

Some other research have focused on {low-rate attacks}: In early research \cite{Macia_Fernandez_2009}, a mathematical model for the LDoS attacks against application servers was developed, which facilitates performance evaluation through its correlation with attack configuration parameters and network-victim dynamics. This model was validated by comparing performance metrics with those obtained through simulation and used to make recommendation for developing defense techniques against the Low-Rate DoS attack.

In order to detect LDoS attack flows, Reference \cite{wu2016} exploited the changes in network traffic characteristics using Multi-Fractal Detrended Fluctuation Analysis (MF-DFA) algorithm. This algorithm was tested via simulations and on a test-bed. Reference~\cite{tang2021detection} developed self-adaptive density-based spatial clustering of applications with noise (SADBSCAN). This algorithm identifies clusters in multi-density datasets, enabling the grouping of network traffic based on low-rate DoS attack characteristics. Subsequently, cosine similarity is employed to ascertain the presence of low-rate DoS attacks within these traffic groups. Reference \cite{aiello2014online} also developed an LDoS detection algorithm that analyzes not packet traffic but spectral features of traffic over time windows. Moreover, Reference \cite{reed2021resource} focused on detecting \emph{Slowloris} attacks in resource constrained IoT networks. To this end, it developed a lightweight intrusion detection framework that captures attack footprints analyzing network events in two stages. 

In order to identify LDoS attacks, the most recent research utilized Machine Learning (ML), especially deep learning, models: Reference~\cite{liu2020detection} proposed multi-feature fusion approach utilizing  K-Nearest Neighbor (KNN) classifier and operating through three features related to the router queue and traffic fluctuations. The multi-feature fusion approach was evaluated on a test-bed. Reference~\cite{cheng2020machine} utilized a machine learning algorithm to analyse traffic packages and identify signatures of attack with low performance overhead. In \cite{xu2021low}, a hybrid neural network architecture, combining 1-Dimensional Convolutional Neural Network (CNN) and Gated Recurrent Unit, was developed. This hybrid architecture was tested on HTTP traffic through experimental work achieving slightly above 98\% accuracy. Reference~\cite{ilango2021, ilango2022feedforward} focused on LDoS attacks targetting Software Defined Network (SDN) of IoT devices and developed FeedForward–Convolutional Neural Network (FFCNN) based anomaly detection scheme. The performance of FFCNN was evaluated on ``CIC DoS 2017'' dataset and compared against well-known ML models, including Multi-Layer Perceptron (MLP), Support Vector Machine (SVM), and Random Forest (RF).

{While the existing low-rate detection models often rely on static feature fusion or reactive classification of current traffic patterns, this paper distinguishes itself by focusing on the temporal evolution of network performance. In contrast with traditional LDoS detection that treats each time window as an independent classification task, our approach utilizes predictive modeling to establish a dynamic baseline of expected network behavior, identifying attacks through the discrepancy between forecast and observed performance metrics.}

\subsection{{Predictive Intrusion Detection}}




{While traditional Intrusion Detection Systems (IDS) focus on identifying attacks as they occur, predictive intelligence leverages historical data and real-time traffic analysis to forecast and mitigate threats before they compromise system integrity }\cite{vengathattil-advanced}{. Recent research highlights the efficiency of learning algorithms in identifying zero-day attacks and insider threats by uncovering patterns that escape signature-based systems }\cite{vengathattil-advanced}{. Several studies have introduced architectural innovations to enhance this predictive accuracy. For instance, to address the challenge of imbalanced datasets in IoT networks, Reference} \cite{maoudj2025deep}{ proposed a hierarchical Deep Learning (DL) approach utilizing a two-step prediction process where a primary model identifies potential attack traffic and a secondary model specifies the exact attack type.}

{In the context of critical sectors like banking, Reference }\cite{akhi2025enhancing}{ demonstrated the superiority of ensemble models over individual algorithms by integrating diverse architectures, including Deep Neural Networks, Random Forest, and Gradient Boosting, to predict threats across heterogeneous datasets. Similarly, addressing the vulnerability of critical network links, Reference }\cite{huang2025defending}{ introduced LinkDam, a framework that tracks and predicts adversary interest to identify potential victim links before an attack reaches its peak, enabling early activation of load balancing. Furthermore, for Industry 4.0 environments, Reference }\cite{al2024situation}{ developed a predictive framework that prioritizes threats based on the specific nature of the attack within Wireless Sensor Networks (WSNs), using a combination of Decision Trees, MLP, and Autoencoders.}

{These advancements underscore the efficacy of predictive intelligence in modern cybersecurity. This paper differentiates from the existing work focused on predictive intrusion detection by focusing on the temporal trends of QoS metrics. By transitioning from simple attack classification to a predictive discrepancy analysis, where the system forecasts expected normal behavior and flags deviations as malicious footprints, the proposed IDQS framework provides a lightweight yet robust solution tailored for the specific dynamics of Low-Rate DoS attacks.}

\subsection{{QoS-Aware Security Modeling}}
{Recent advancements in network security research highlight a reciprocal link between network performance metrics and the identification of low-rate attacks. Unlike volumetric DDoS attacks that aim for total service disruption, LDoS attacks specifically target QoS by exploiting protocol-level adaptive mechanisms, such as TCP congestion control and the three-way handshake of the MQTT protocol }\cite{rios2022detection, vedula2021detection}{. These low and slow maneuvers are strategically designed to diminish throughput and increase latency while remaining below the detection thresholds of traditional volume-based security systems }\cite{tang2022adms}.

{In Reference~}\cite{hajimaghsoodi2022rad} and \cite{ccakmak2024impact}{, empirical evidence from SDN and LTE cellular contexts suggests that the degradation of performance parameters, specifically packet loss, jitter, and end-to-end delay, serves as a primary indicator of an ongoing LDoS or slow-rate incursion. Furthermore, contemporary mitigation frameworks such as PeakSAX and ADMS utilize real-time monitoring of these QoS fluctuations to distinguish between legitimate network congestion and malicious performance degradation} \cite{tang2022adms}{. By evaluating detection latency and accuracy as performance benchmarks, Reference~}\cite{hernandez2025real}{ emphasizes that maintaining high QoS is a vital defensive posture against sophisticated, low-rate security threats.}

References~\cite{de2025detection} and \cite{chiu2021stealthy} {examine scenarios where QoS metrics function as the primary victims in stealthy attacks such as Reduction of Quality (RoQ) and DeGrading of Service (DGoS). These studies illustrate how attackers prioritize subtle performance erosion over total service cessation, utilizing link manipulation to evade detection by network tomography }\cite{chiu2021stealthy, yousef2024unveiling}. For instance, Reference~\cite{wu2022research}{ analyzes how these maneuvers manifest in power communication networks (PCNs) as increased transmission delays and packet loss, which undermine the reliability of critical infrastructure without triggering binary connectivity alarms.}

{Consequently, Reference~}\cite{moreira2025cyberattacks} {repurposed statistical quality management tools, such as multivariate control charts, to detect abnormal traffic patterns in cyber-physical systems by treating security breaches as performance defects. In specialized domains like the Internet of Medical Things (IoMT) and vehicular networks, References~}\cite{nguyen2021dependability} and \cite{baldini2022detection}{ identify timing-related observables as critical indicators of operational integrity and security status. Furthermore, Reference~}\cite{yakin2021security} and \cite{sawas2023real}{ illustrate this trend by leveraging Signal-to-Noise Ratio (SNR), data rates, and power spectrum analysis within the NeMo-IoT and LSTM-based prediction models to identify deviations from established baseline behaviors.}

{Despite this growing recognition, a significant research gap remains in the transition from passive observation to proactive, predictive security modeling. Majority of existing frameworks rely on static threshold-based detection or reactive anomaly classification, which struggle to differentiate between legitimate temporal drifts and the subtle signatures of LDoS attacks. This paper addresses this gap by introducing Intrusion Detection via QoS Prediction, in short IDQS, a system-level holistic design that shifts the paradigm from simple classification to predictive discrepancy analysis. In contrast with the traditional absolute-threshold models, the proposed IDQS utilizes a specialized Recurrent Trend Predictive QoS Predictor to anticipate expected normal behavior, coupled with a Pairwise Decision Model that acts as a comparator layer to identify malicious footprints through subtle forecasting errors. By modeling the underlying temporal trends of QoS metrics, this approach provides a robust defense mechanism tailored for the resource-constrained environments of IoT gateways, where early and accurate detection is critical for maintaining infrastructure resilience.
}




\section{Holistic System for Intrusion Detection via Quality-of-Service Prediction}\label{sec:Method}

\begin{figure*}[t!]
	\centering
	\includegraphics[scale=0.475]{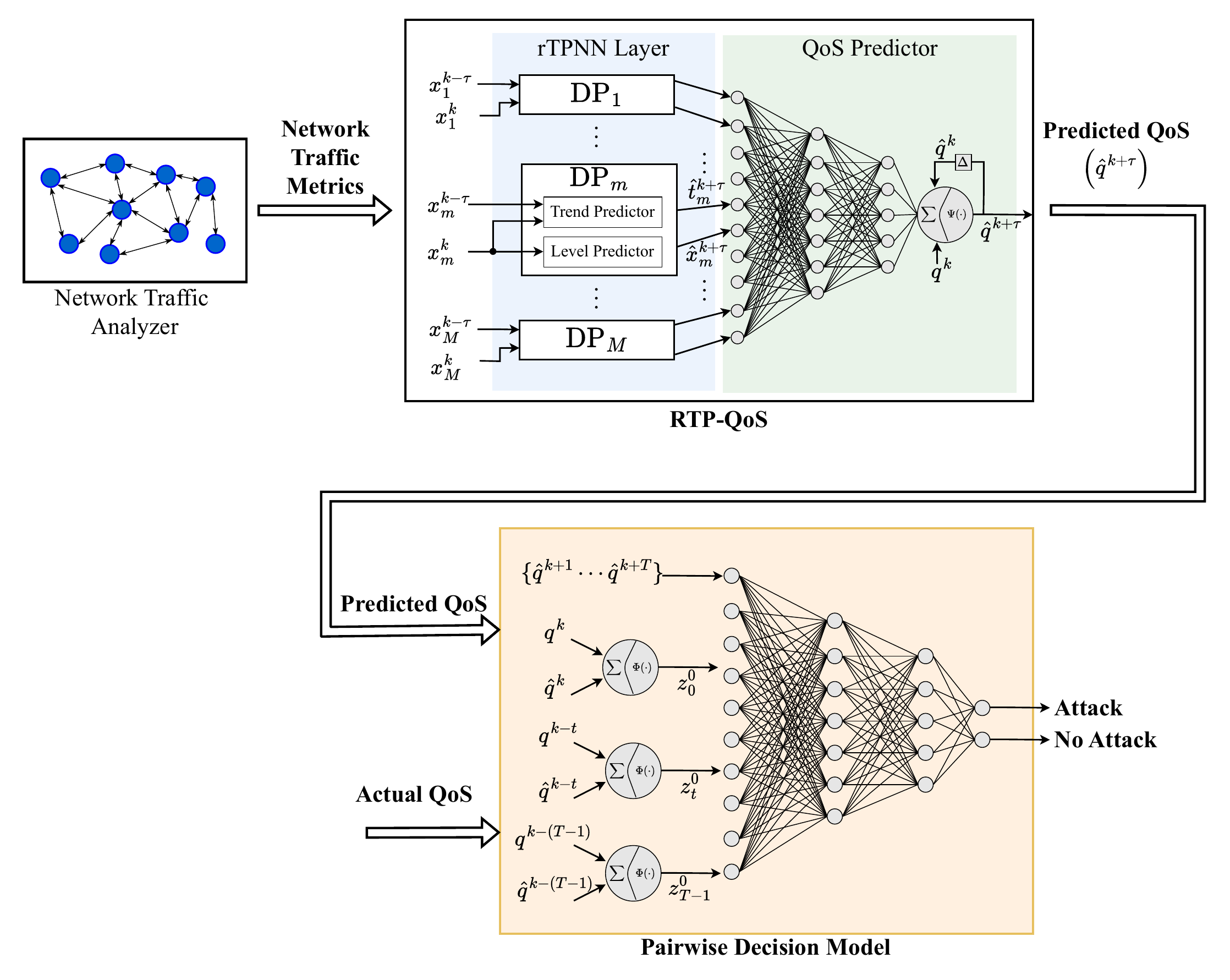}
	\caption{Structure of the proposed Intrusion Detection via QoS Prediction System (IDQS)}\label{fig:rTPNN_LDOS}
\end{figure*}

This section presents the proposed holistic system for Intrusion Detection via QoS Prediction, called IDQS, which is displayed in Figure~\ref{fig:rTPNN_LDOS}. The IDQS identifies the advanced Low-Rate DoS (LDoS) attacks via the Pairwise Decision Model (PDM) based on the QoS predicted by an enhanced rTPNN architecture, called Recurrent Trend Predictive QoS Predictor (RTP-QoS). Due to the design of the PDM architecture and the internal structure of RTP-QoS, which predicts the upcoming trends and values of traffic features, it is expected to detect Low-Rate DDoS attacks in their very early stages. 

The proposed IDQS operates over the traffic flows. For each traffic flow, $M$ predefined features are extracted to support prediction and detection by the RTP-QoS and PDM models, where the value of each feature $m \in \{1, \dots, M\}$ is denoted by $x_m^k$. After each traffic flow, indexed by $k$, is completed, the RTP-QoS neural network architecture predicts the QoS for the future traffic flow $k+\tau$, denoted by $\hat{q}^{k+\tau}$, based on the collection of current and the past network traffic flow features, i.e. $\{(x_m^{k}, \, x_{m}^{k - 1})\}_{m \in \{1, \dots, M\}}$. Then, the predicted QoS $\hat{q}^{k+\tau}$ is fed into the PDM architecture for detecting a possible LDoS attack effecting or that is predicted to effect the network performance. 
One should note that throughput (in bytes per second) is the QoS metric that is sought to be predicted in this paper, whereas the proposed methodology does not depend on any particular selection of the QoS metric.



\subsection{Network Traffic Features}\label{sec:Features}

In order to obtain a lightweight and practically applicable system, the features extracted from the network traffic are defined as
\begin{itemize}
	\item Average number of packets transmitted per second,
	\item Average number of bytes per packet,
	\item Average number of flows per second, and
	\item Average life of a traffic flow.
\end{itemize}
These features are selected to make computations easily, protect privacy of the traffic content, and provide information to forecast future QoS. In addition to these features, within the proposed RTP-QoS neural network architecture, the past measurements of QoS are also used. Subsequently, each traffic feature is normalized by dividing its value by the maximum value of that feature across all experiments considered. While the normalization is effective for shortening the learning time of the IDQS components, it is not a mandatory operation and does not affect the model performance.

One should note that the features considered in this paper are not part of the proposed IDQS and does not effect its design. Therefore, based on the properties of the applications or datasets for which IDQS shall be tested, the sets of features can be extended, shortened, or redefined.

\subsection{Recurrent Trend Predictive QoS Predictor (RTP-QoS)}

Within the proposed IDQS, first, the network QoS, $\hat{q}^{k+\tau}$, for the upcoming traffic flow $k+\tau$ is predicted via the Recurrent Trend Predictive QoS Predictor, namely RTP-QoS, model. As shown in Figure~\ref{fig:rTPNN_LDOS}, this model is comprised of two main parts: rTPNN layer and QoS Predictor. 

\subsubsection{rTPNN Layer}
The rTPNN layer within RTP-QoS is a critically modified version of the characteristic layer of the original rTPNN \cite{rTPNN}. Although the original rTPNN extracts the current trend and level of the sequential input data, our modified version of the rTPNN layer now predicts the upcoming trend and upcoming value of each traffic feature. In this way, the RTP-QoS model, which contains the modified rTPNN layer, foresees the upcoming traffic and capture the relationship between traffic trend and QoS prediction.  

To this end, the modified rTPNN layer is comprised of the ``Level Predictor'' and the ``Trend Predictor'' modules, each of which consists of a recurrent neuron. The objectives of these modules are respectively predicting the value and trend of the associated traffic feature for the future flow. 

At each traffic flow $k$, the Level Predictor module of the modified rTPNN layer predicts the upcoming value of feature $m$ as
\begin{equation}\label{eq:Level}
	\hat{x}_m^{k+\tau} = \beta_{m, 1}^k \, x_m^k + \beta_{m, 2}^k \, x_m^{k-\tau} + \beta_{m, 3}^k \, \hat{x}_m^k , 
\end{equation}
where $\hat{x}_m^{k+\tau}$ denotes the predicted value of feature $m$ for flow $k+\tau$, and $\{ \beta_{m, 1}, \beta_{m, 2}, \beta_{m, 3}\}$ is the set of learnable parameters of the recurrent neuron forming the Level Predictor module. 

The Trend Predictor module of the modified rTPNN layer predicts the trend of feature $m$ for traffic flow $k+\tau$ as 
\begin{equation}\label{eq:Trend}
	\hat{t}_m^{k+\tau} = \alpha_{m, 1}^k \, \bigl( x_m^k - x_m^{k-\tau} \bigr) + \alpha_{m, 2}^k \, \bigl( \hat{x}_m^{k+\tau} - x_m^{k} \bigr) + \alpha_{m, 3}^k \, \hat{t}_m^k , 
\end{equation}
where $\hat{t}_m^{k+\tau}$denotes the predicted value of the trend of feature $m$ for traffic flow $k+\tau$, and $\alpha_{m, 1}$, $\alpha_{m, 2}$, and $\alpha_{m, 3}$ are the learnable parameters of the recurrent neuron forming the Trend Predictor module. In (\ref{eq:Trend}), in addition to the actual feature values taken for consecutive flows $k-\tau$ and $k$, the feature value $\hat{x}_m^{k+\tau}$ predicted by the Level Predictor module and the trend $\hat{t}_m^k$ predicted earlier for the flow $k$ are also used to provide a successful prediction of the upcoming trend.

Finally, the level and trend values predicted for each of the $M$ traffic features for the flow $k+\tau$ are concatenated into a vector $\hat{y}_0^{k+\tau}$:  
\begin{equation}
	\hat{y}_0^{k+\tau} = \Bigl[ \hat{x}_m^{k+\tau}, \, \hat{t}_m^{k+\tau}: ~~ m \in \{1, \dots, M\} \Bigr],
\end{equation}
which is the input of the QoS Predictor part of the RTP-QoS model.

\subsubsection{QoS Predictor} The next part of the RTP-QoS is QoS Predictor, which is --as shown in Figure~\ref{fig:rTPNN_LDOS}-- comprised of three successive fully connected layers and a recurrent output neuron. While the fully connected layers process and merge the predicted trends and levels of all traffic features, the recurrent output neuron provides the final prediction of the network QoS.  

The fully connected layers are feed-forward and comprised of neurons with nonlinear activation functions. Hence, each layer $l \in \{1, \dots, L\}$ simply operates as follows to process the output of the previous layer:  
\begin{equation}
	\hat{y}_l^{k+\tau} \leftarrow \psi \bigl( \, W_l^k \, \hat{y}_{l-1}^{k+\tau} \, \bigr),
\end{equation}
where $y_l^{k+\tau}$ is the output vector of fully connected layer $l$, $W_l^k$ is the input weight matrix of layer $l$ calculated after the completion of flow $k$, and $\psi(\cdot)$ is an element-wise activation function. 

Subsequently, the output of the fully connected layers, denoted by the vector $\hat{y}_L^{k+\tau}$, which is the processed version of the trend and level predictions, is used by the recurrent output neuron to predict the network QoS, denoted by $\hat{q}^{k+\tau}$, for the upcoming traffic flow $k+\tau$. In addition to the $\hat{y}_L^{k+\tau}$ vector, the output neuron process the earlier prediction and observation of the network QoS for the current flow $k$. These inputs, one of which is the neuron's own output, serve as error-correcting feedback for the QoS Predictor, helping to improve prediction accuracy over time. Accordingly, the recurrent output neuron calculates the QoS prediction as
\begin{equation}
	\hat{q}^{k+\tau} = \Psi\bigl( \gamma_1^k \, \hat{y}^{k+\tau} +  \gamma_2^k \, q^k + \gamma_3^k \, \hat{q}^{k+\tau} \bigr),
\end{equation}
where $\gamma_1^k$, $\gamma_2^k$ and $\gamma_3^k$ are the learnable parameters, i.e. connection weights of the output neuron, and $\Psi(\cdot)$ is an element-wise activation function. 

\subsection{Pairwise Decision Model for Low-Rate DoS Detection}\label{sec:PDM}
After the transmission of the traffic flow $k+\tau$ is completed, the proposed IDQS decides if the flow carries a signature of low-rate attack or the QoS predictions provide an insight towards experiencing the impact of such attack. To this end, we develop the Pairwise Decision Model (PDM), which is a lightweight neural network architecture. The PDM is designed to capture the pairwise relationships between the predicted and the actual QoS values observed in the past and combine these relationships with the future forecast of QoS to identify signature of a low-rate attack on the traffic. 

The PDM architecture proposed has two sets of inputs: 
\begin{itemize}
	\item The first input is the set of the actual-forecast QoS pairs obtained in the current and the past timeslots: $\{(q^{k-t}, \, \hat{q}^{k-t} )\}_{t \in \{0,\dots,T-1\}}$.
	
	\item The other input is the set of forecast QoS values for the upcoming $T$ time slots: $\{\hat{q}^{k+t}\}_{t \in \{1,\dots,T\}}$
\end{itemize}

As shown in Figure~\ref{fig:rTPNN_LDOS}, the first layer is designed to process the each pair of the actual and the forecast QoS. To this end, this layer, namely pairwise layer, is comprised of $T+1$ neurons, each of which, indexed by $t \in \{0, \dots, T-1\}$ associated to the corresponding pair of inputs, ie. $(q^{k-t}, \, \hat{q}^{k-t} )$. The main purpose of each neuron $t$ in the pairwise layer is to learn which relationship between the actual and the forecast QoS is the most import for detecting a low-rate attack. Accordingly, the output of neuron $t$ in the pairwise layer is calculated as
\begin{equation}
	z^0_t = \phi(\omega_t^1 q^{k-t} + \omega_t^2 \hat{q}^{k-t}),
\end{equation}
where $\omega_t^1$ and $\omega_t^2$ are the connection weights respectively for the inputs of $q^{k-t}$ and $\hat{q}^{k-t}$. In addition, $\phi(\cdot)$ is a nonlinear activation function. 

Then, the collection of neuron outputs at the pairwise layer constructs the output of this layer: 
\begin{equation}
	Z^0 = \bigl[ z^0_0, z^0_1, \dots, z^0_{T-1} \bigr] 
\end{equation}
Subsequently, the output of the pairwise layer is concatenated with the second set of inputs, ie. the set of forecast QoS values $\{\hat{q}^{k+t}\}_{t \in \{1,\dots,T\}}$, in order to process the past and the current pairwise QoS relations with the forecast QoS for the future traffic. The resulting vector is processed by the first fully connected layer of PDM: 
\begin{equation}
	Z^1 = \Phi\bigl(\, \Omega^1 \, [Z^0, \hat{q}^{k+1}, \dots, \hat{q}^{k+T}]^\intercal \, \bigr),
\end{equation}
where $\Omega^1$ is the input weight matrix for the first fully connected layer of PDM, and $\Phi(\cdot)$ is a nonlinear activation function. After the first fully connected layer, for each of the following fully connected layers, the output is calculated as
\begin{equation}
	Z^h = \Phi\bigl(\, \Omega^h \, Z^{h-1} \, \bigr), \qquad  h \in \{2, \dots, H-1\}.
\end{equation}
The output layer of the PDM architecture is a softmax layer, which is comprised of two neurons corresponding to ``No Attack'' and ``Attack'' cases, and whose output is 
\begin{equation}
	Z^H = softmax\bigl(\, \Omega^h \, Z^{H-1} \, \bigr).
\end{equation}
The resulting decision on the attack is taken as
\begin{equation}
	d^k = \arg \max\bigl(Z^H\bigr), 
\end{equation}
where the element indexed by $0$ refers to the case ``No Attack'', and the element indexed by $1$ refers to the case ``Attack''.

\section{Results} \label{sec:Results}

This section now presents the performance evaluation results for predicting QoS and detecting LDoS. The performance of the proposed RTP-QoS is comprehensively evaluated for QoS prediction and compared against the well-known ML models. The proposed architecture is finally tested for detecting LDoS attacks. This architecture is also evaluated under two decision models, namely the proposed Pairwise Decision Model (PDM) and Fully Connected Decision Model (FCDM). 

The experimental work and the performance evaluation in this paper are conducted using an online dataset, called ``SDN-SlowRate-DDoS'', \cite{dataset}. This dataset contains both normal and attack traffic recordings collected through experiments with varying parameters of attack duration, attack rate, and number of attackers. In this paper, the numerical results are presented for the experiments lasting 4000 seconds as those can provide the sufficiently long data for both training test.


\subsection{QoS Prediction Performance}

The QoS prediction performance of the RTP-QoS model is evaluated through the experiment that contains only the normal traffic. To this end, the experiment data is divided into training and test sets of size 80\% and 20\%, respectively. In order to protect the time series characteristics of the data, the training and test sets are extracted in timely order.

\subsubsection{Analyses of Varying Time Step Predictions}

First, the impact of the time step $\tau$ on the QoS prediction performance of RTP-QoS is analysed. To this end, the RTP-QoS model is trained and tested distinctly for each value of $\tau$ in $\{$10~s, 50~s, 100~s, 150~s, 200~s, 250~s, 300~s, 350~s, 400~s$\}$. This analysis provides a valuable insight towards the robustness of RTP-QoS against the selection of the value of $\tau$ and also enables us to set the $\tau$ for the rest of the performance analyses. 

The QoS prediction performance of the RTP-QoS model is given in Table~\ref{tab:tau_values} with respect to Mean Squared Error (MSE), Mean Absolute Error (MAE), and Normalized Symmetric Mean Absolute Error (N-SMAPE). Note that the Symmetric Mean Absolute Error (SMAPE) is normalized dividing by 2, so that N-SMAPE has value in range 0\%-100\%. The results in this table reveal that the QoS prediction performance of the proposed architecture is considerably stable and robust against the selection of $\tau$ as it achieves less than 2\% error in N-SMAPE for all values of $\tau$ tested. Meanwhile, the performance comparison across different values of $\tau$ remains consistent for all metrics.

\begin{table}[h!]
	\centering
	\normalsize
	\renewcommand{\arraystretch}{1.15}
	\setlength{\tabcolsep}{6pt}
	\caption{The performance of the RTP-QoS model for QoS prediction under normal network traffic for varying values of $\tau$}
	\begin{tabular}{|c||c|c|c|}
		\hline
		\multirow{3}{*}{\textbf{\begin{tabular}[c]{@{}c@{}}TIME\\STEP\end{tabular}}} & \multicolumn{3}{c|}{\textbf{Test Performance}} \\ \cline{2-4} 
		& {\textbf{\begin{tabular}[c]{@{}c@{}}MSE\\ ($\times 10^{-3}$)\end{tabular} }} & {\textbf{\begin{tabular}[c]{@{}c@{}}MAE\\ ($\times 10^{-3}$)\end{tabular} }} & {\textbf{N-SMAPE}} \\ \hline
		10 s & 1.51 & 33.88 & 1.83\% \\ \hline
		50 s & 0.51 & 16.44 & 0.87\% \\ \hline
		100 s & 0.76 & 20.96 & 1.11\% \\ \hline
		150 s & 0.57 & 17.99 & 0.95\% \\ \hline
		200 s & 0.45 & 16.18 & 0.85\% \\ \hline
		250 s & 0.95 & 26.32 & 1.39\% \\ \hline
		300 s & 0.49 & 17.13 & 0.89\% \\ \hline
		350 s & 0.65 & 20.98 & 1.11\% \\ \hline
		400 s & 0.82 & 23.45 & 1.24\% \\ \hline
	\end{tabular}
	\label{tab:tau_values}
\end{table}

The fluctuating performance of RTP-QoS across different values of $\tau$ does not lead to a definitive conclusion; however, certain values of $\tau$, in particular 50~s, 150~s, 200~s, and 300~s, stand out with N-SMAPE of less than $1\%$. Based on this analysis, we set $\tau=50$ for the rest of the performance evaluation in this section. Although $\tau=200~s$ provides the best prediction, it results in four times greater data loss during analysis (corresponding to a longer cold-start in real-life implementations) compared to $\tau=50~s$. In particular, for $\tau = 200~s$, the first 200 seconds --approximately $5\%$-- of the data are used solely for providing model inputs, while the last 200 seconds are dedicated exclusively to generating prediction outputs. Consequently, around $10\%$ of the data is effectively lost when evaluating samples. This is only $2.5\%$ for $\tau=50~s$.

\subsubsection{Predicting QoS During Normal Traffic}

\begin{figure}[h!]
	\centering
	\includegraphics[scale=0.325]{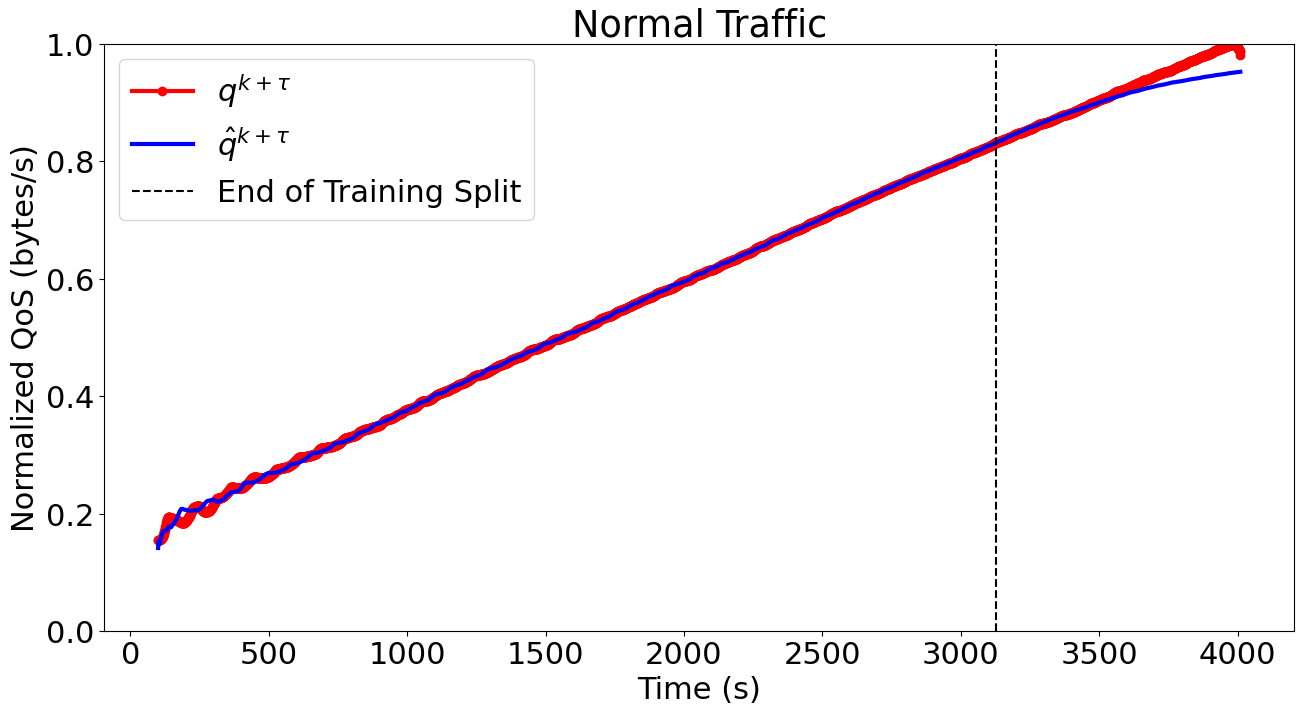}
	\caption{QoS predicted by the modified rTPNN and the actual QoS under normal network traffic}\label{fig:QoS_normal}
\end{figure}

The actual and predicted QoS values for $\tau=50~s$ are presented in Figure~\ref{fig:QoS_normal}. These results demonstrate that the RTP-QoS effectively learns from the training set. Moreover, the predicted QoS values closely match the actual values, accurately capturing the trends in the test set. Comparable greater error is observed only at the peaks of the actual values, although the predicted QoS still reflects the trend and general behaviour of the actual QoS.

Furthermore, Table~\ref{tab:QoS_normal_metrics_comparison} presents the average prediction error of RTP-QoS and its comparison against the well known Long Short-Term Memory (LSTM), Multilayer Perceptron (MLP), and Nonlinear Perceptron (NP) models for training and test sets. 

\begin{table}[h!]
	\centering
	\normalsize
	\renewcommand{\arraystretch}{1.15}
	\setlength{\tabcolsep}{6pt}
	\caption{Comparison of the RTP-QoS model against the well-known benchmark models for QoS prediction under normal network traffic}
	\begin{tabular}{|c|c|c|c|}
		\hline
		\multicolumn{4}{|c|}{\textbf{Training Performance}}  \\ \hline
		\textbf{Models} & {\textbf{\begin{tabular}[c]{@{}c@{}}MSE\\ ($\times 10^{-3}$)\end{tabular} }} & {\textbf{\begin{tabular}[c]{@{}c@{}}MAE\\ ($\times 10^{-3}$)\end{tabular} }} & {\textbf{N-SMAPE}} \\ \hline
		RTP-QoS & 0.016 & 2.29 & 0.39\% \\ \hline
		LSTM & 0.025 & 3.44 & 0.47\%	 \\ \hline
		MLP & 0.032 & 4.09 & 0.56\% \\ \hline
		NP & 0.084 & 7.22 & 0.86\% \\ \hline \hline
		
		\multicolumn{4}{|c|}{\textbf{Test Performance}}  \\ \hline
		\textbf{Models} & {\textbf{\begin{tabular}[c]{@{}c@{}}MSE\\ ($\times 10^{-3}$)\end{tabular} }} & {\textbf{\begin{tabular}[c]{@{}c@{}}MAE\\ ($\times 10^{-3}$)\end{tabular} }} & {\textbf{N-SMAPE}} \\ \hline
		RTP-QoS & 0.51 & 16.44 & 0.87\% \\ \hline
		LSTM & 3.463 & 52.75 & 2.88\% \\ \hline
		MLP & 4.43 & 58.99 & 3.23\% \\ \hline
		NP & 4.22 & 61.05 & 3.36\% \\ \hline
	\end{tabular}
	\label{tab:QoS_normal_metrics_comparison}
\end{table}

\begin{figure*}[t!]
	\centering
	\includegraphics[scale=0.32]{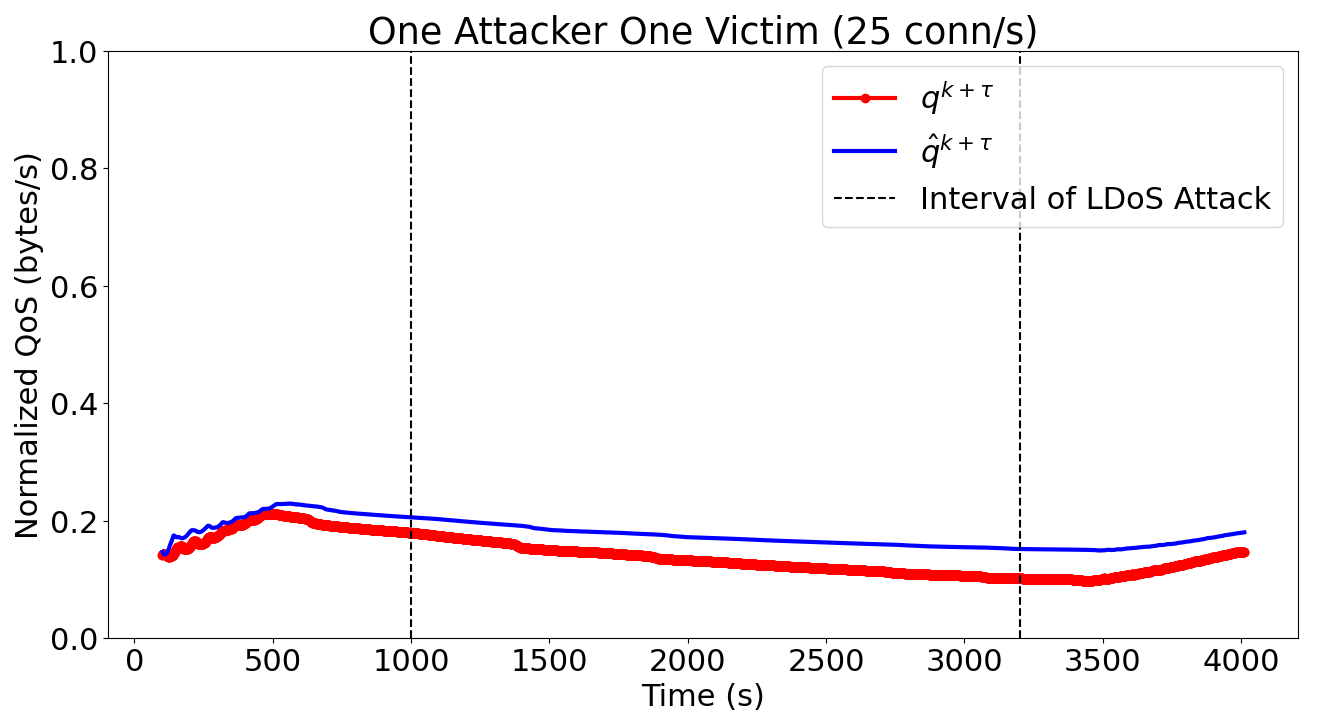}~
	\includegraphics[scale=0.32]{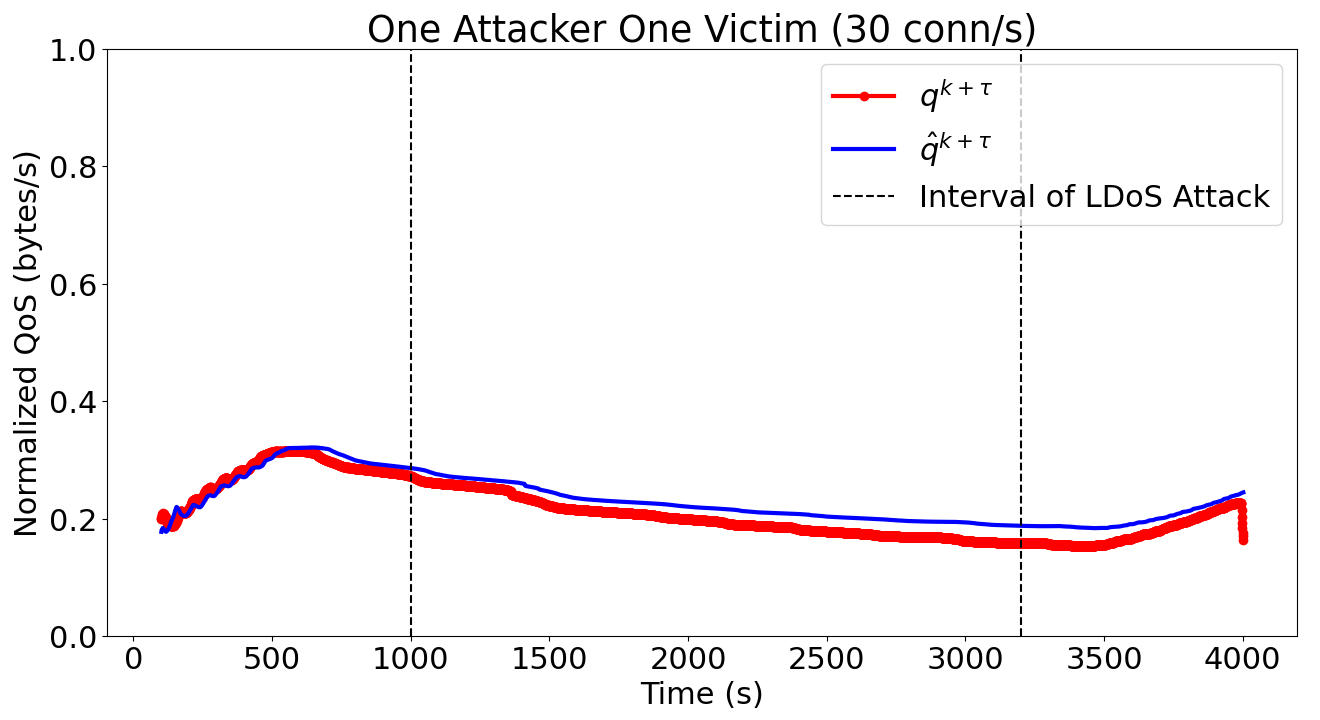}\\\vspace{0.1cm}
	\includegraphics[scale=0.32]{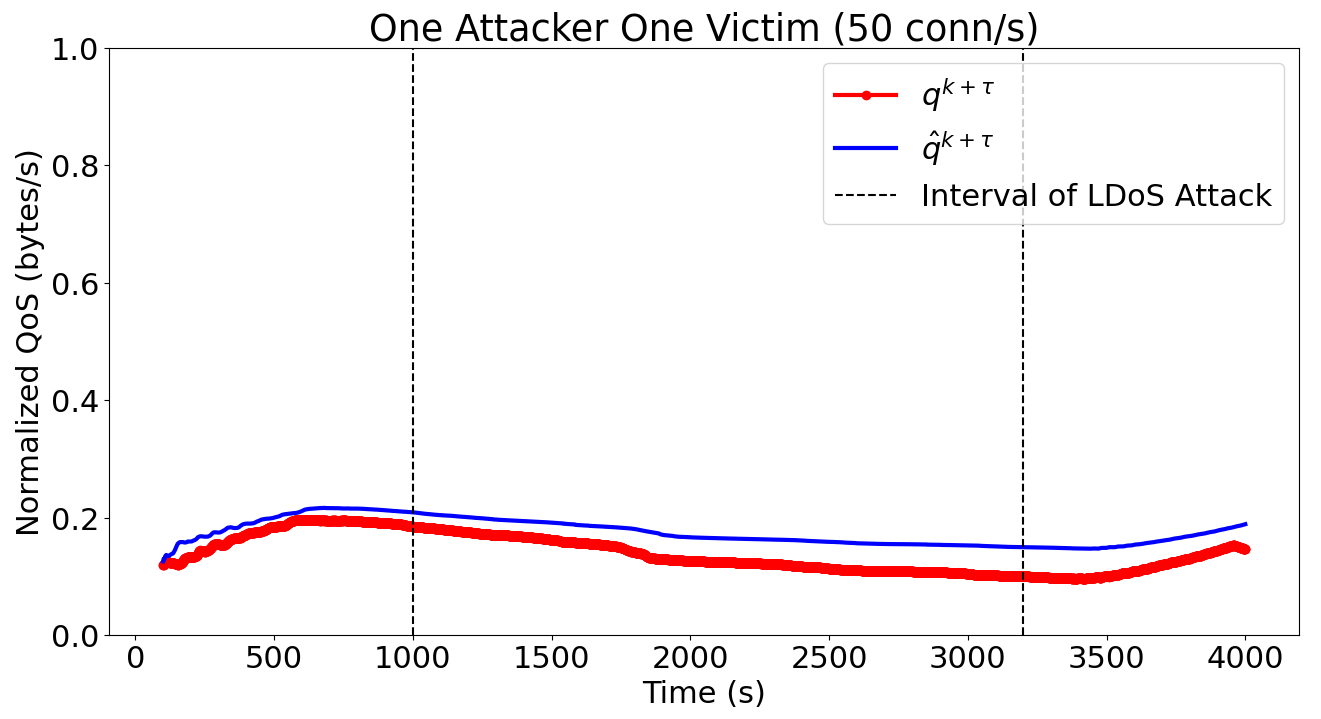}~
	\includegraphics[scale=0.32]{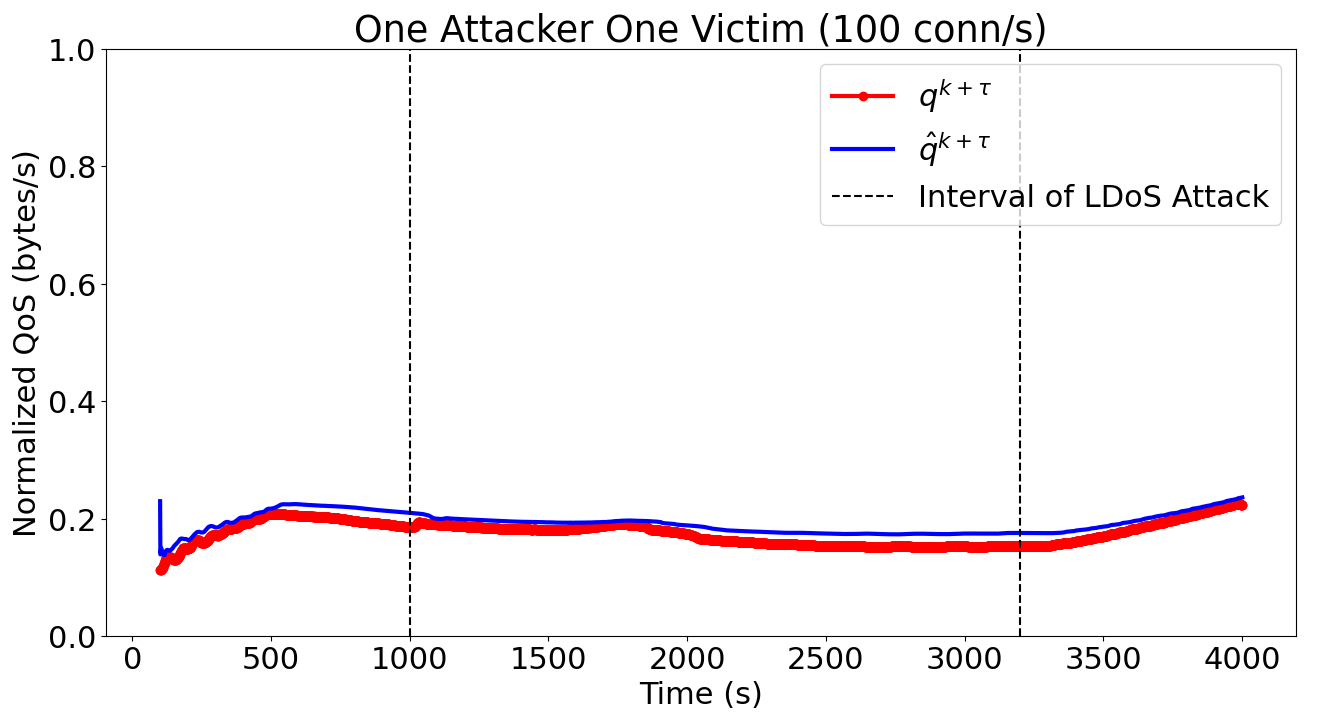}
	\caption{The predicted and the actual QoS values under LDoS attacks with a single attacker creating 25 connections per second (conn/s), 30 conn/s, 50 conn/s, and 100 conn/s, respectively, from top to bottom.}
	\label{fig:QoS_oneAttacker}
\end{figure*}

The results in Table~\ref{tab:QoS_normal_metrics_comparison} show that RTP-QoS achieves the lowest prediction error among all models in terms of MSE, MAE and N-SMAPE, on both training and test sets. The RTP-QoS model attains an N-SMAPE value of $0.39$ on the training set and $0.87$ on the test set. While all evaluated models achieve training errors below $1\%$ N-SMAPE, RTP-QoS is the only model to sustain prediction accuracy with an N-SMAPE below $1\%$ on the test set, demonstrating performance at least three times superior to that of other evaluated models.

{This superior performance on the test set (N-SMAPE < 1\%) is critical for the operational viability of RTP-QoS. By establishing such a high-fidelity baseline for normal traffic, IDQS ensures that even minor deviations caused by stealthy LDoS maneuvers can be statistically isolated. Unlike traditional models where a 3\% error might mask a low-rate attack, the precision of RTP-QoS allows the system to treat the forecast-observed discrepancy as a clean signal for anomaly detection rather than model noise.}

\subsection{Predicting QoS During LDoS Attack}

In order to analyse the behaviour of RTP-QoS during low-rate attacks and observe the possible impact of QoS prediction in attack detection, we now present the prediction results of RTP-QoS during the LDoS attacks available in the dataset.

Figure~\ref{fig:QoS_oneAttacker} displays the predictions obtained by the RTP-QoS model for the LDoS attack with one victim and one attacker. The prediction results are taken for attack rates corresponding to 25, 30, 50, and 100 connections per second. The results in this figure show that the predictions of RTP-QoS are successfully aligned with the actual QoS for the normal traffic until the LDoS attack starts. In addition, the results in Figure~\ref{fig:QoS_oneAttacker} show that the actual QoS drops below the predicted QoS over time after the attack begins. One should note that the results obtained for different attack scenarios provided almost the same insights. 

\begin{figure}[h!]
	\centering
	\includegraphics[scale=0.23]{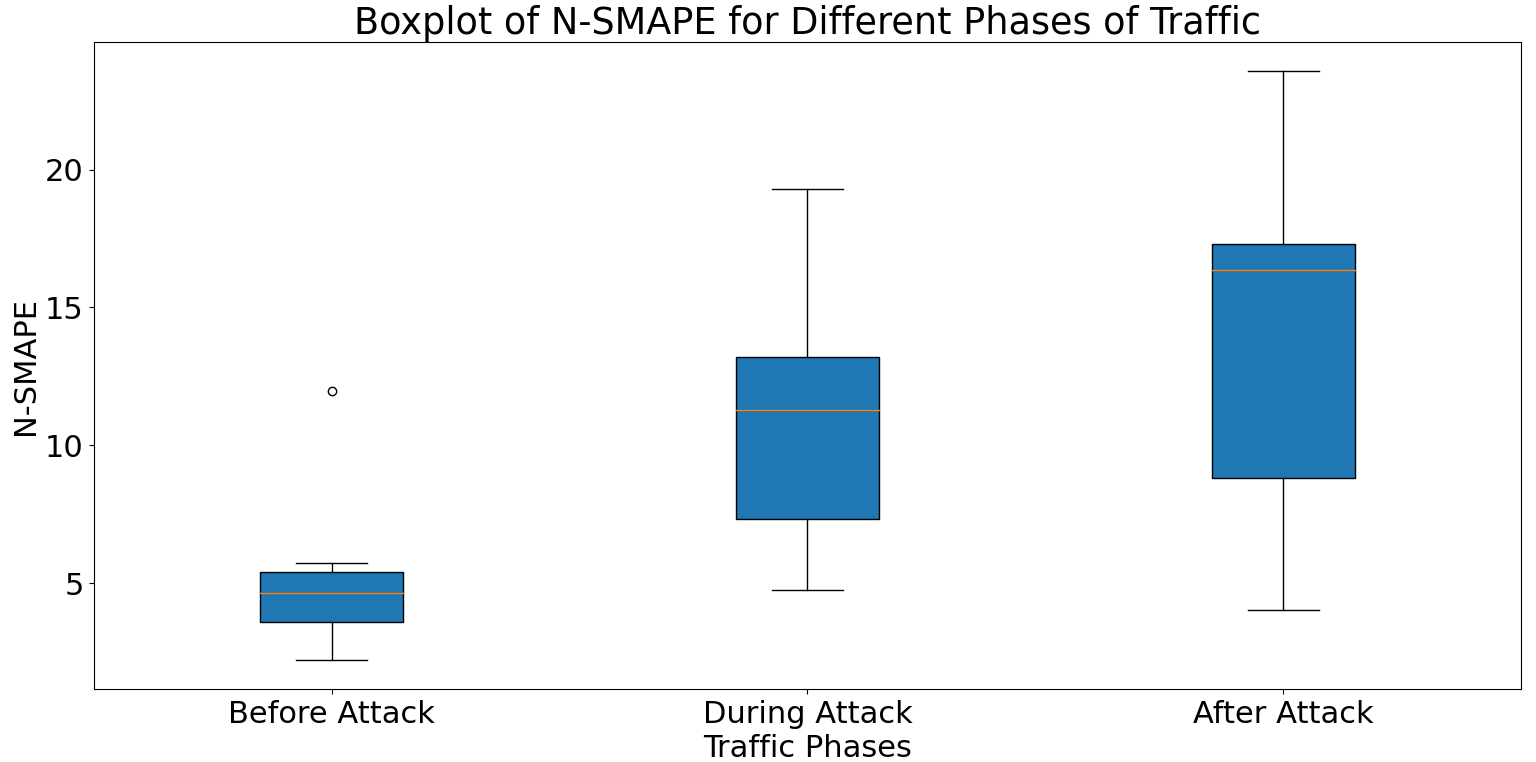}
	\caption{Boxplot of N-SMAPE calculated for three different phases of network traffic and presented among all experiments files containing an attack}\label{fig:error_boxplot}
\end{figure}

\begin{table*}[t!]
	\centering
	\renewcommand{\arraystretch}{1.5}
	\normalsize
	\caption{Performance of the Proposed Approach on Test Experiments During 5-Fold CV}
	\begin{tabular}{|c||c|c|c|c|}
		\hline
		\textbf{Experiment} & \textbf{Accuracy} & \textbf{Precision} & \textbf{Recall} & \textbf{F1-Score} \\ \hline\hline
		One Attacker One Victim (25 conn/s) & 0.85 & 0.82 & 0.96 & 0.88 \\ \hline
		One Attacker One Victim (30 conn/s) & 0.59 & 0.67 & 0.57 & 0.62 \\ \hline
		One Attacker One Victim (50 conn/s) & 0.91 & 0.86 & 1.00 & 0.92 \\ \hline
		One Attacker One Victim (100 conn/s) & 0.80 & 0.79 & 0.88 & 0.83 \\ \hline\hline
		Four Attackers One Victim (10 conn/s - 15 conn/s) & 0.79 & 0.73 & 1.00 & 0.85 \\ \hline
		Four Attackers One Victim (20 conn/s - 30 conn/s) & 0.83 & 0.77 & 1.00 & 0.87 \\ \hline
		Four Attackers Two Victims (10 conn/s - 15 conn/s) & 0.84 & 0.88 & 0.83 & 0.86 \\ \hline
		Four Attackers Two Victims (20 conn/s - 30 conn/s) & 0.81 & 0.82 & 0.86 & 0.84 \\ \hline\hline
		Three Attackers Three Victims (30 conn/s) & 0.58 & 0.58 & 1.00 & 0.73 \\ \hline
	\end{tabular}
	\label{tab:detection_cv}
\end{table*}

Moreover, Figure~\ref{fig:error_boxplot} indicates the relationship between prediction error (measured in N-SMAPE) and the initiation of a low-rate attack. For the vast majority of experimental data, we observe a significant increase in prediction error after the initiation of an attack. We also observe that the error continues to increase for some time after the attack stops. On the other hand, for the case of ``One Attacker One Victim (100 conn/s)'', the N-SMAPE remains almost the same through all phases of traffic. The errors measured for that experiment are $4.65\%$, $4.76\%$, and $4.03\%$ respectively for ``Before Attack'', ``During Attack'', and ``After Attack'' phases. 

{The increasing divergence between predicted and actual QoS observed in Figure~}\ref{fig:QoS_oneAttacker}{ represents the adversarial gap. Since the RTP-QoS model is an idealized representation of healthy network behavior, this gap serves as a quantifiable metric of the progressive impact of the attack. The persistence of high error even after the attack concludes (as seen in Figure~}\ref{fig:error_boxplot}{) highlights the slow-recovery nature of LDoS attacks, which exhaust resources in a way that lingers beyond the active packet flow, a signature that IDQS is uniquely equipped to track.}

Based on the results shown in Figure~\ref{fig:QoS_oneAttacker} and Figure~\ref{fig:error_boxplot} collectively, relying solely on prediction error for LDoS detection may lead to false alarms and misdetections. In particular, accurately identifying the initiation and termination of attacks is expected to be highly challenging due to the slow and gradual nature of their impact. It can also be inferred (especially from Figure~\ref{fig:QoS_oneAttacker}) that achieving a sufficiently large error margin to effectively distinguish malicious or abnormal activities from the normal traffic requires a considerably long detection interval --if indeed feasible-- after the initiation of the attack. Therefore, smart and proactive approaches, e.g. Pairwise Decision Model presented in this paper, are expected to perform better than straightforward approaches.

\subsection{Low-Rate Attack Detection via PDM}

Throughout the evaluation of the low-rate attack detection performance of our proposed architecture, we perform 5-fold cross-validation (CV). Table~\ref{tab:detection_cv} displays the average performance of our approach over the CV folds with respect to Accuracy, Precision, Recall, and F1-Score.

The results in Table~\ref{tab:detection_cv} show that the proposed approach achieves an accuracy of over 0.79 for the majority (7 out of 9) attack cases. For these cases, our approach has achieved a considerably high F1-score of over 0.83. In addition, Recall remains high and Precision exhibits variability, suggesting that the model tends to minimize false negatives but encounters challenges in consistently minimizing false positives.

\setcounter{figure}{5}
\begin{figure*}[t!]
	\centering
	\includegraphics[width=\linewidth, height=9cm]{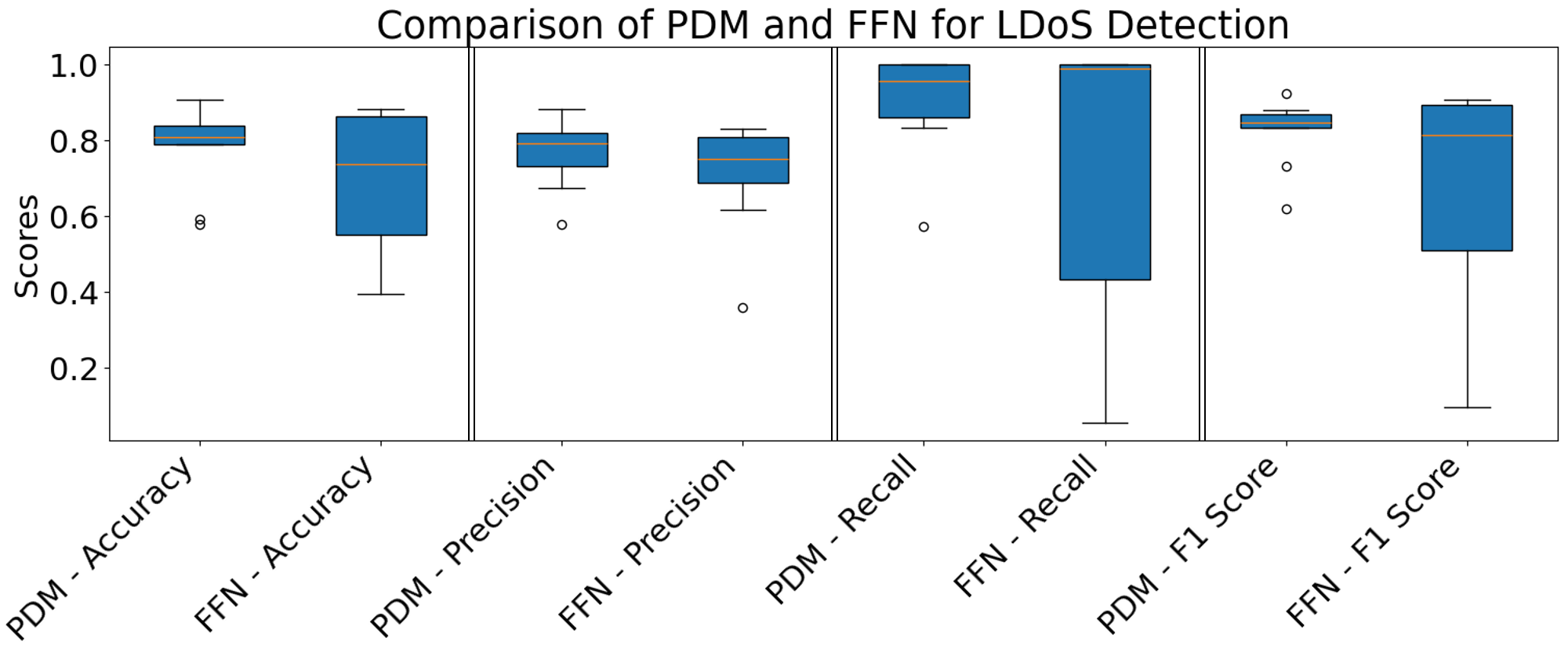}
	\caption{The boxplot of the performances of the PDM and FFN over the CV folds for each of Accuracy, Precision, Recall, and F1 Score}\label{fig:LDoS_comparison}
\end{figure*}

On the other hand, the proposed approach can achieve only 0.59 and 0.58 accuracy for the cases of ``One Attacker One Victim (30 conn/s)'' and ``Three Attackers Three Victims''. This is observed to be due to the lack of training data similar to the test case within the fold of CV. In order to validate this observation, for the cases of ``One Attacker One Victim (30 conn/s)'', we compare the performance under CV against that when the model is trained using only the data for the cases with One Attacker (excluding the 30 conn/s case).

\setcounter{figure}{4}
\begin{figure}[h!]
	\centering
	\includegraphics[scale=0.3]{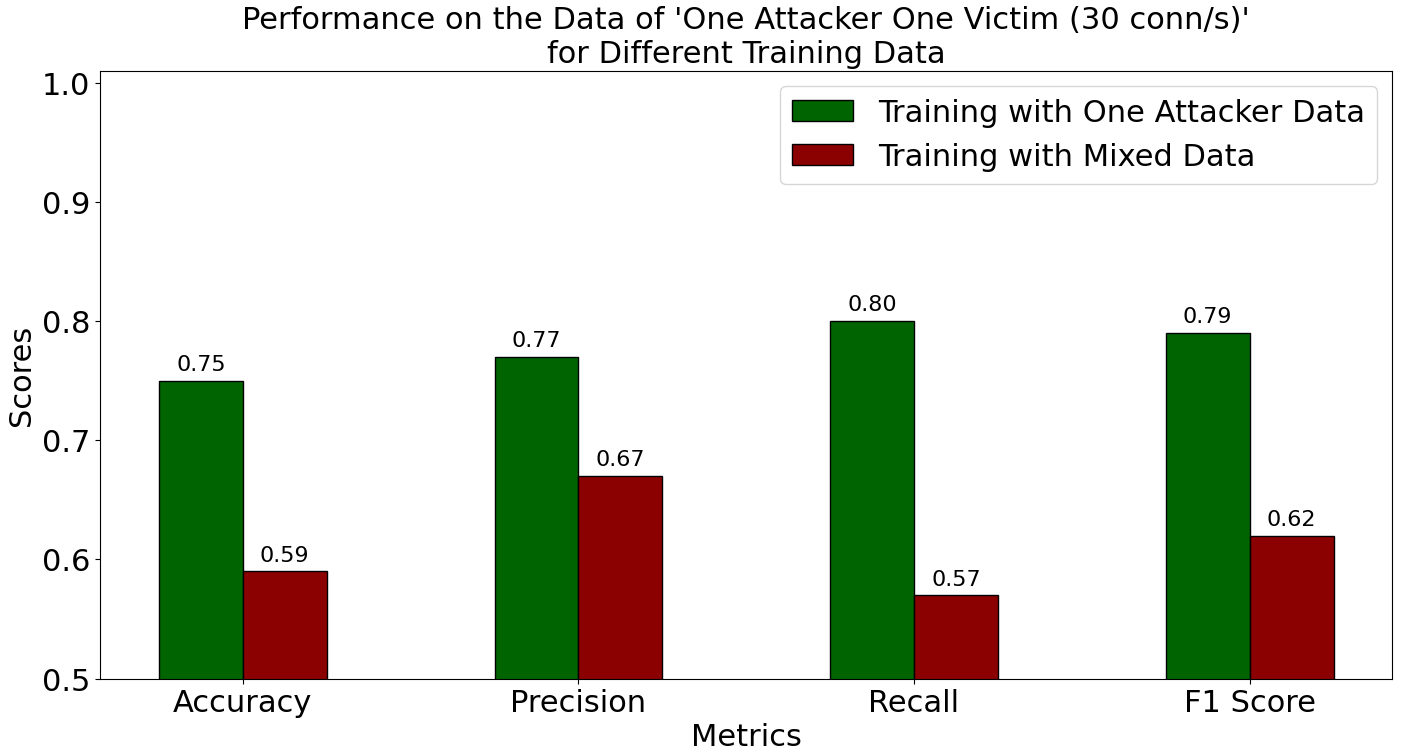}
	\caption{Performance of the proposed approach for the cases of ``One Attacker One Victim (30 conn/s)'' under two different training cases}\label{fig:TrainingDataExp2}
\end{figure}
\setcounter{figure}{6}

On the other hand, the proposed approach achieves relatively low accuracy, in particular 0.59 and 0.58, for the cases of ``One Attacker One Victim (30 conn/s)'' and ``Three Attackers Three Victims'', respectively. This is observed to be due to the lack of training data similar to the test case within the CV folds where these cases are used for testing. In order to validate this observation, we examine the performance for the ``One Attacker One Victim (30 conn/s)'' case by comparing the results under CV with those obtained when the model is trained exclusively on data from other ``One Attacker'' cases, excluding the 30 conn/s instance. 
Figure~\ref{fig:TrainingDataExp2} displays the results of the examination. The results in this figure show that training the model with other ``One Attacker'' cases significantly improves the performance in terms of both detecting the normal and the attack traffic.

\subsubsection{Comparison Against the Benchmark}

We further compare the performance of the proposed Pairwise Decision Model (PDM) against the well-known Feed-Forward Network (FFN). In order to present the detailed performance analysis and comparison for PDM and FFN, Figure~\ref{fig:LDoS_comparison} displays the boxplot of the performances of the PDM and FFN over the CV folds for each metric. 

The results in this figure show that the proposed pairwise structure outperforms the fully connected network across all metrics. For these metrics, one may see that the interquartile range of the PDM's performance is considerably narrow, indicating that the PDM provides robust and reliable detection across different training and test splits in CV. 

\begin{figure*}[h!]
	\centering
	\includegraphics[scale=0.32]{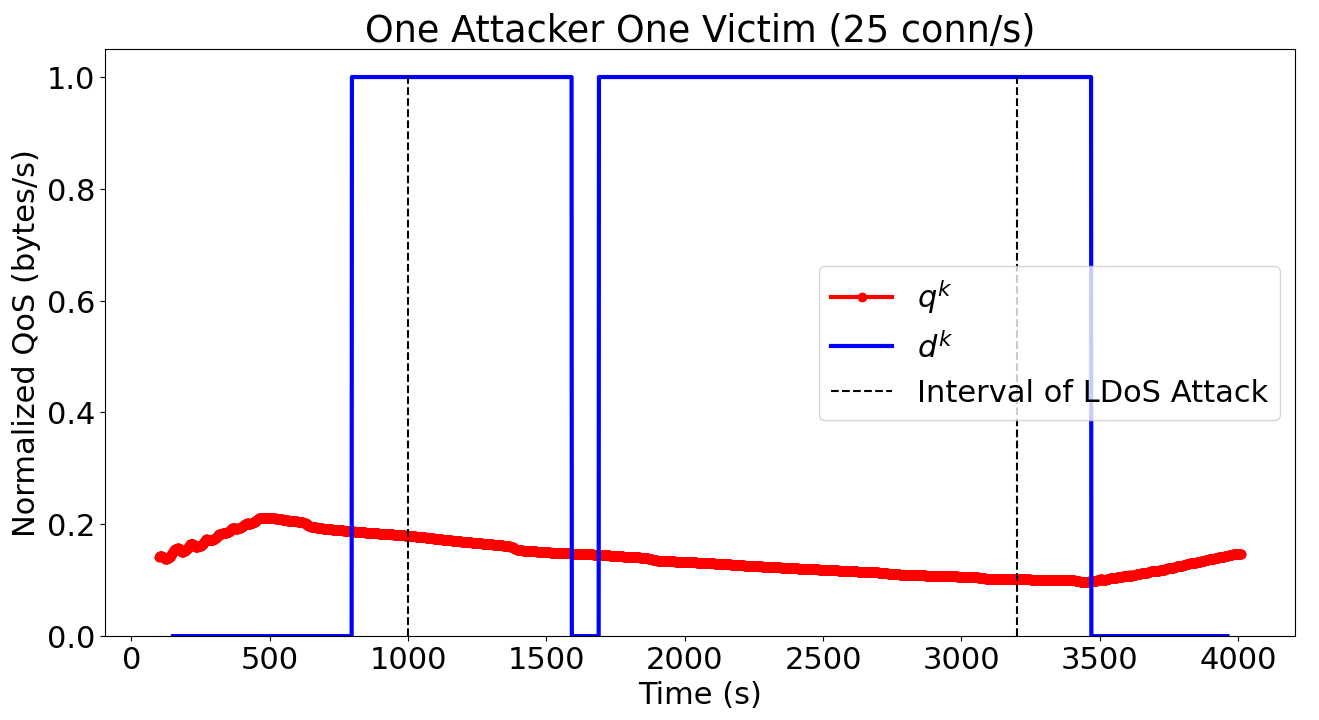}~
	\includegraphics[scale=0.32]{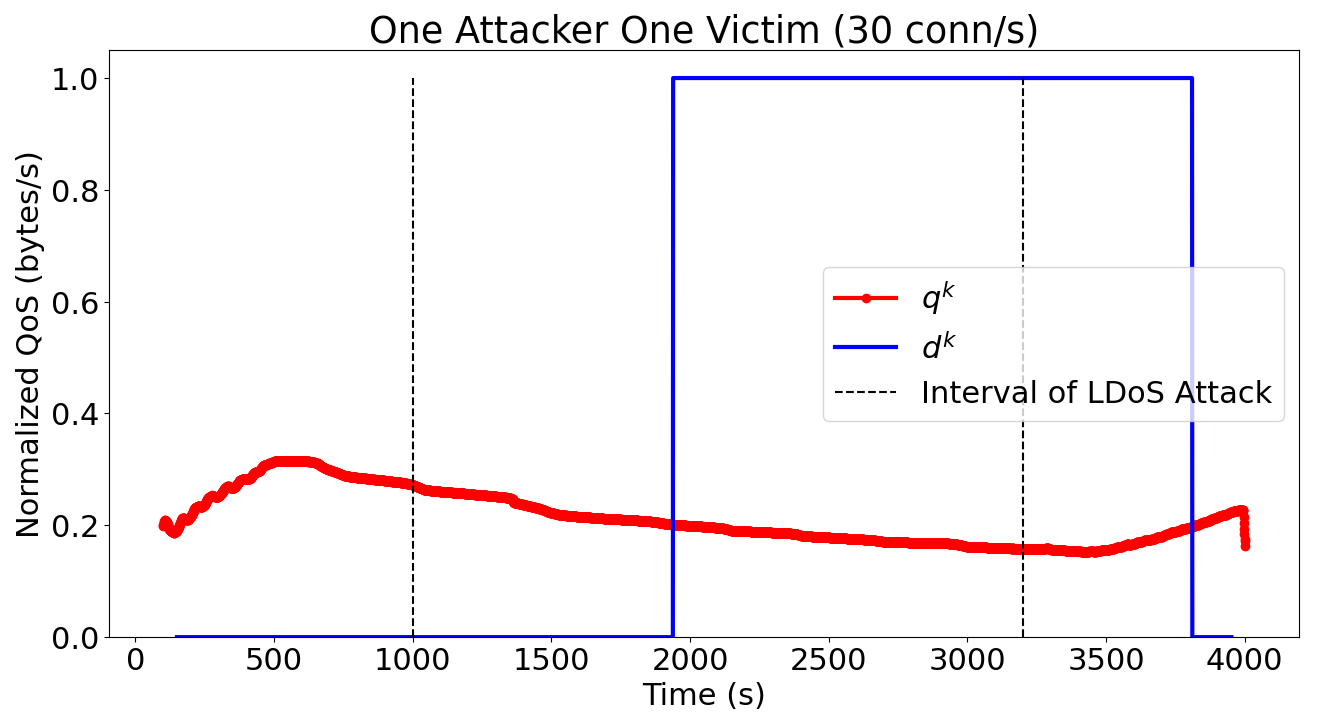}\\\vspace{0.1cm}
	\includegraphics[scale=0.32]{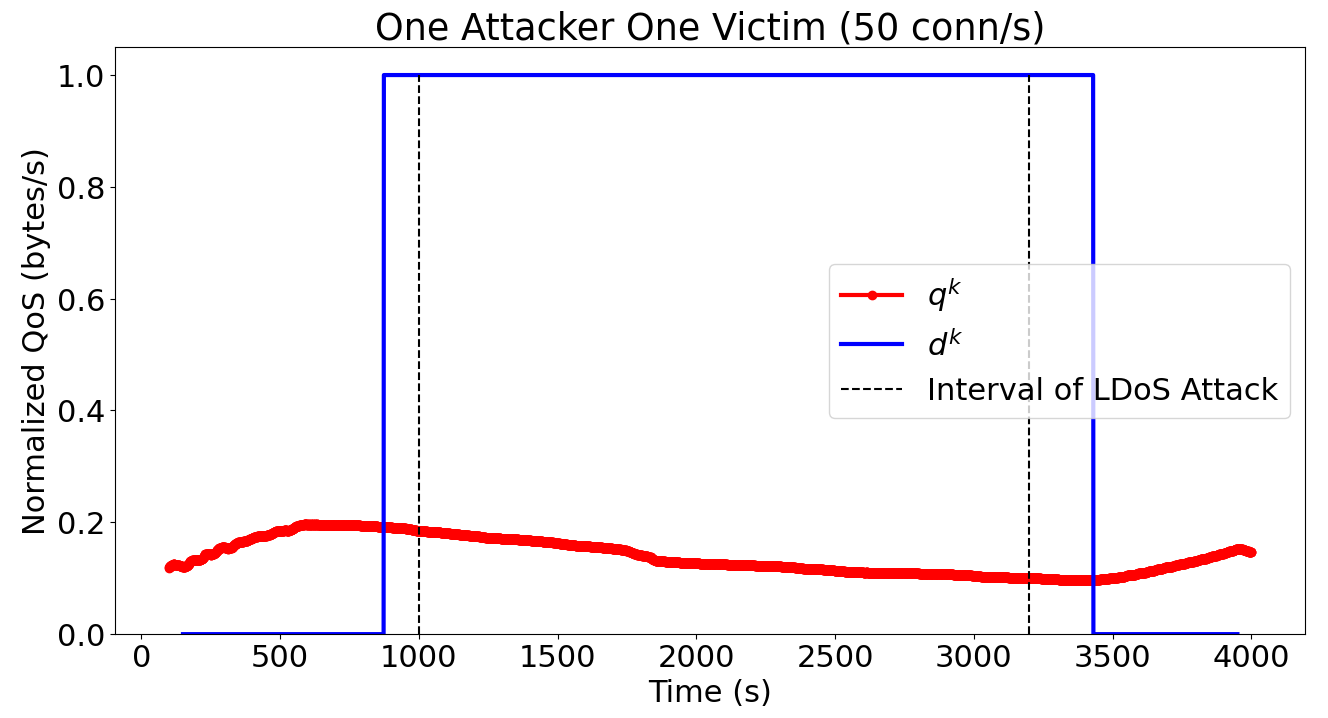}~
	\includegraphics[scale=0.32]{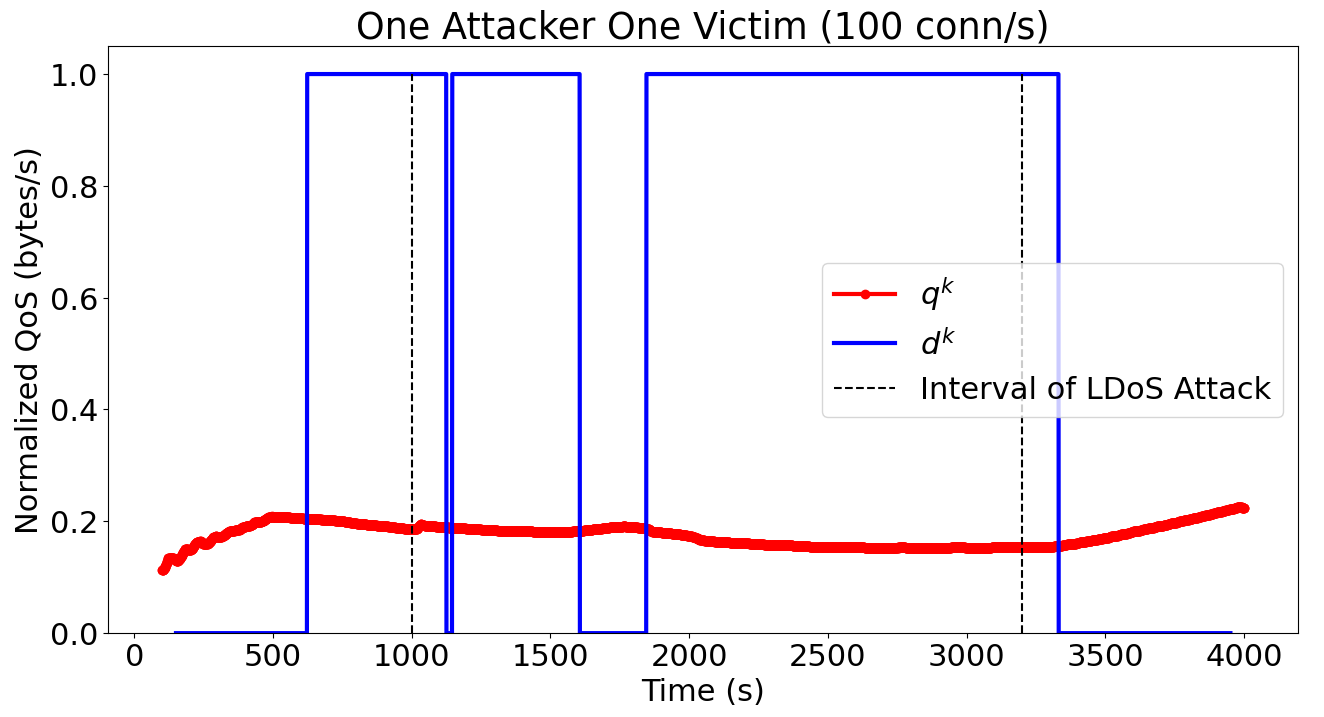}
	\caption{The actual QoS values and corresponding detection results under LDoS attacks with a single attacker creating 25 connections per second (conn/s), 30 conn/s, 50 conn/s, and 100 conn/s, respectively, from top to bottom.}
	\label{fig:detection_oneAttacker}
\end{figure*}

In greater detail, the results displayed in Figure~\ref{fig:LDoS_comparison} exhibits the following comparisons for each evaluation metric:
\begin{itemize}
	\item \textbf{Accuracy}: The PDM demonstrates higher median accuracy with lower variability compared to FFN, which exhibits a wider spread and the minimum accuracy of $0.4$. This indicates that PDM consistently performs better across folds.
	\item \textbf{Precision}: The PDM has higher precision than FFN for all folds of CV, indicating its superiority in correctly identifying positive instances with minimal false positives.  
	\item \textbf{Recall}: Although the median recall of FFN is slightly higher than that of PDM, FFN suffers from significant fluctuation. Indeed, the best recall that FFN obtained is $1$ while the worst is $0.06$. On the other hand, recall of the proposed PDM is above $0.84$ for all CV folds, except one outlier with recall of $0.57$. This underscores PDM's ability to better identify true positives, making less false negatives, which is critical in LDoS detection.
	\item \textbf{F1 Score}: The F1 Score, being the harmonic mean of precision and recall, again highlights the trustable and robust performance of the PDM, which has a very narrow interquartile range and the worst outlier with F1 Score of $0.62$. On the other hand, wider range and lower values of FFN again highlight its inconsistent detection capability.
\end{itemize}
The cumulative results show that the proposed PDM consistently outperforms FFN, across all evaluation metrics, with significantly more stable and reliable performance across cross-validation folds.

Furthermore, we present the actual decisions made by the proposed approach for ``One Attacker One Victim'' cases in Figure~\ref{fig:detection_oneAttacker}. The results in this figure show that the proposed approach successfully captures the LDoS attack traffic for three of the four cases. For the cases with high accuracy, the proposed approach captures the changes before the labeled interval of attack duration. On the other hand, Figure~\ref{fig:detection_oneAttacker} (top-right) shows that the low accuracy obtained for the case ``One Attacker One Victim (30 conn/s)'' is because the propose approach captures the impact of attack considerably late. In particular, for this case, the attack is detected almost 1000 seconds after it started.  

{The performance gap between PDM and FFN clarifies the architectural necessity of the pairwise layer. While a standard FFN attempts to classify traffic based on static snapshots, the PDM operates as a differential sensor. It focuses on the relationship between expectation (prediction) and reality (measurement). This allows it to ignore global traffic surges that affect both values equally while triggering alarms on the subtle asymmetric drifts caused by protocol-level exploitation.}


\subsection{{Performance on the CIC-IDS2017 Dataset}}

{In order to further validate the generalization capability and robust performance of the proposed IDQS, we conduct an additional evaluation using the well-known CIC-IDS2017 dataset }\cite{CCIDS2017_dataset}{. Since IDQS is specifically designed for the continuous time-series analysis of low-rate attacks, our experiments focus exclusively on the traffic recorded on Wednesday. This specific subset is selected as it is the only part of the dataset that contains representative LDoS scenarios, namely DoS Slowloris (between 9:47 and 10:10 a.m.) and DoS Slowhttptest (between 10:14 and 10:35 a.m.), which align with the target attack profiles of our framework. The evaluation is performed with a time step of $\tau = 50$, consistent with the earlier analysis in this paper.}

\begin{figure}[h!]
	\centering
	\includegraphics[scale=0.115]{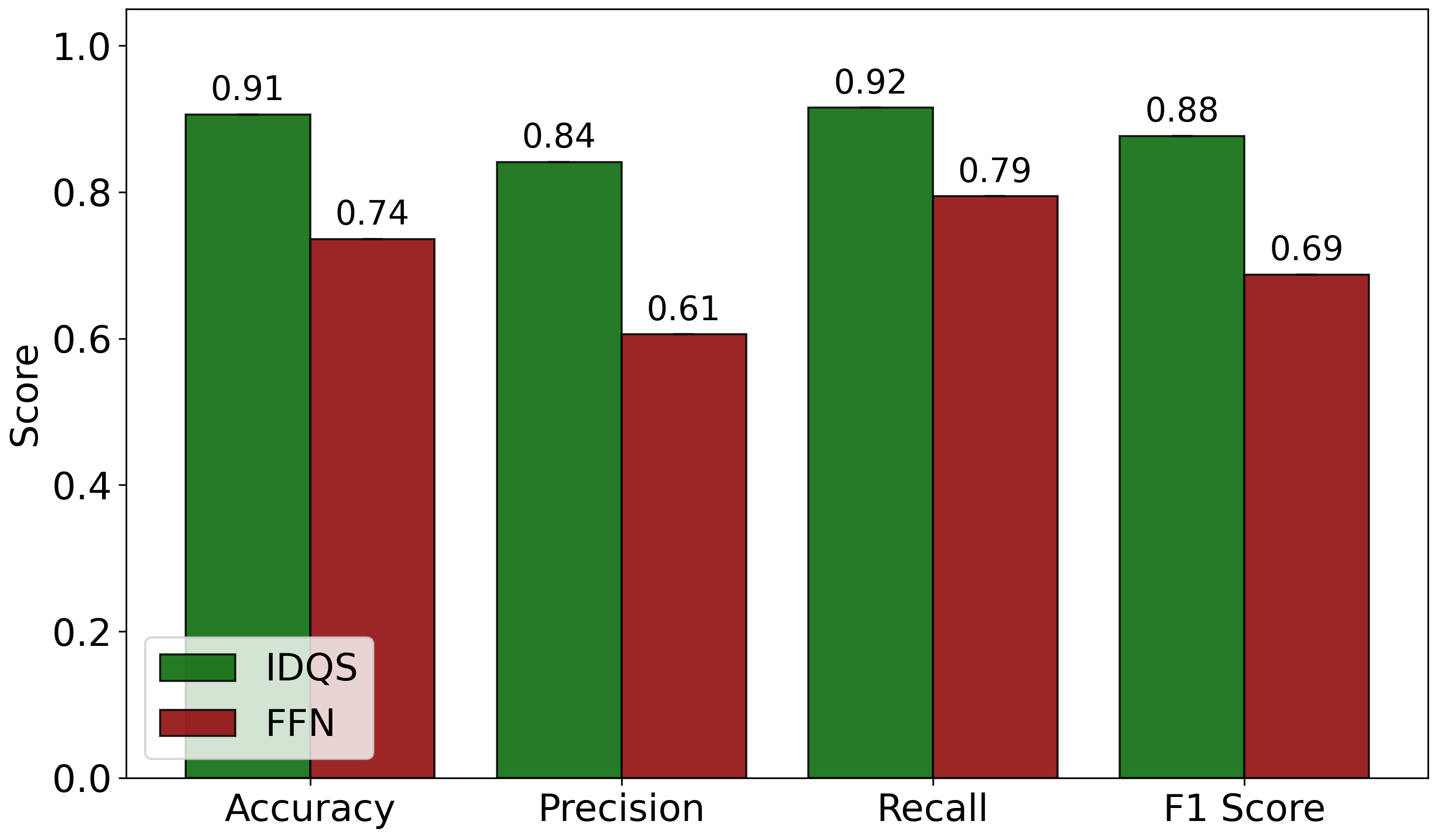}
	\caption{Performance of the proposed approach on the CIC-IDS2017 Wednesday dataset for low-rate attack detection}\label{fig:cic_ids}
\end{figure}

{Figure~}\ref{fig:cic_ids}{ presents the comparative performance of the proposed IDQS (utilizing RTP-QoS and PDM) against the benchmark Feed-Forward Network (FFN). FFN used as a benchmark is comprised of five hidden layers and a single output layer, where the hidden layers respectively contain 128, 64, 32, 16, and 8 neurons.}

{The results displayed in this figure demonstrate that the proposed architecture significantly outperforms the FFN across all evaluated metrics. Specifically, IDQS achieves an accuracy of 91\%, which is a significant improvement over the 74\% accuracy of the benchmark model. Furthermore, the high recall of 0.92 underscores the effectiveness of the proposed system in identifying subtle LDoS signatures, while maintaining a robust F1 Score of 0.88. These findings confirm that by shifting from static classification to predictive discrepancy analysis, IDQS maintains high detection reliability even when deployed on heterogeneous datasets containing varied low-rate attack vectors.}

\subsection{Training and Inference Time}\label{sec:CompTimeResults}

\begin{figure}[h!]
	\centering
	\includegraphics[scale=0.24]{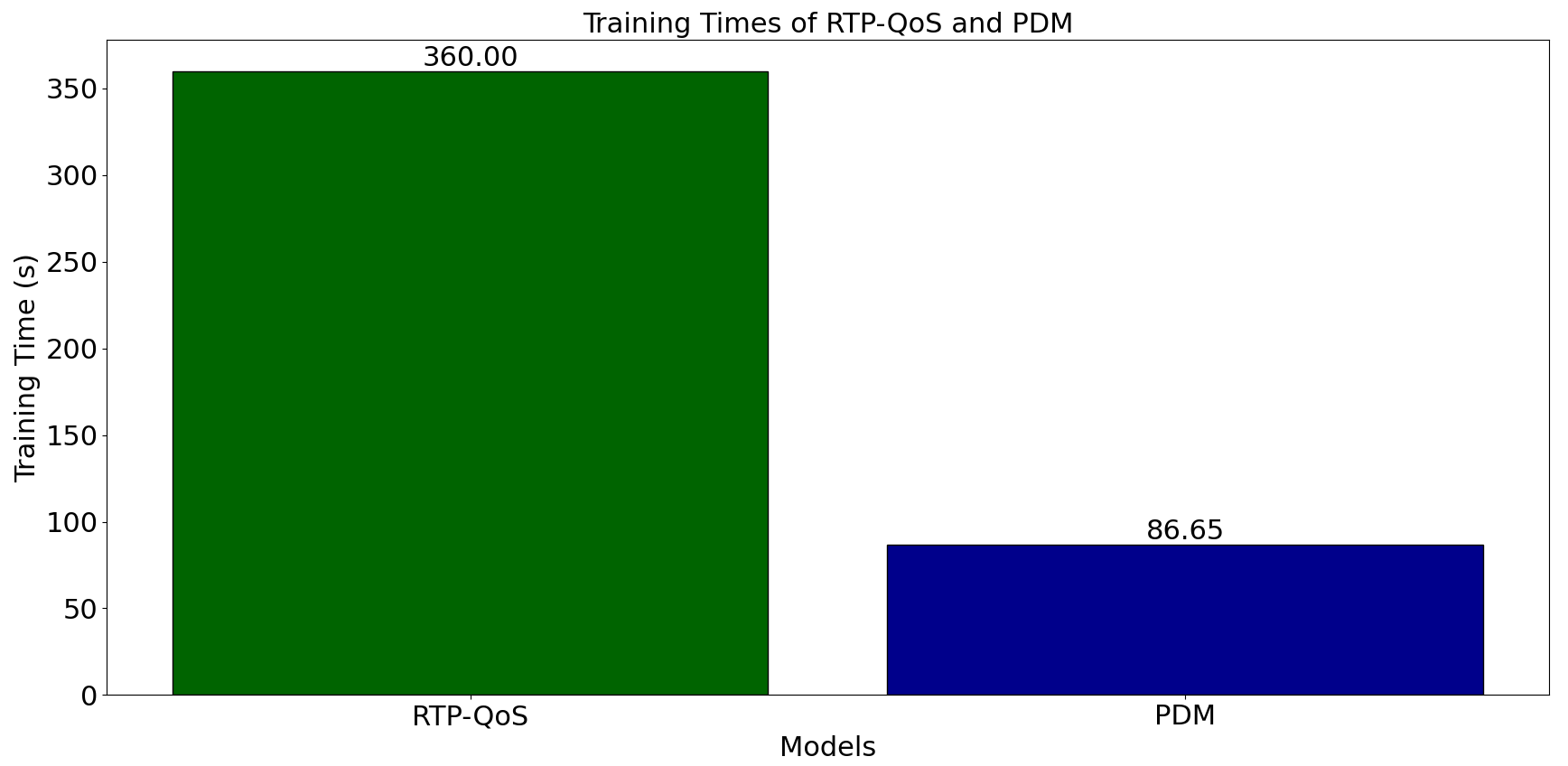}
	\includegraphics[scale=0.24]{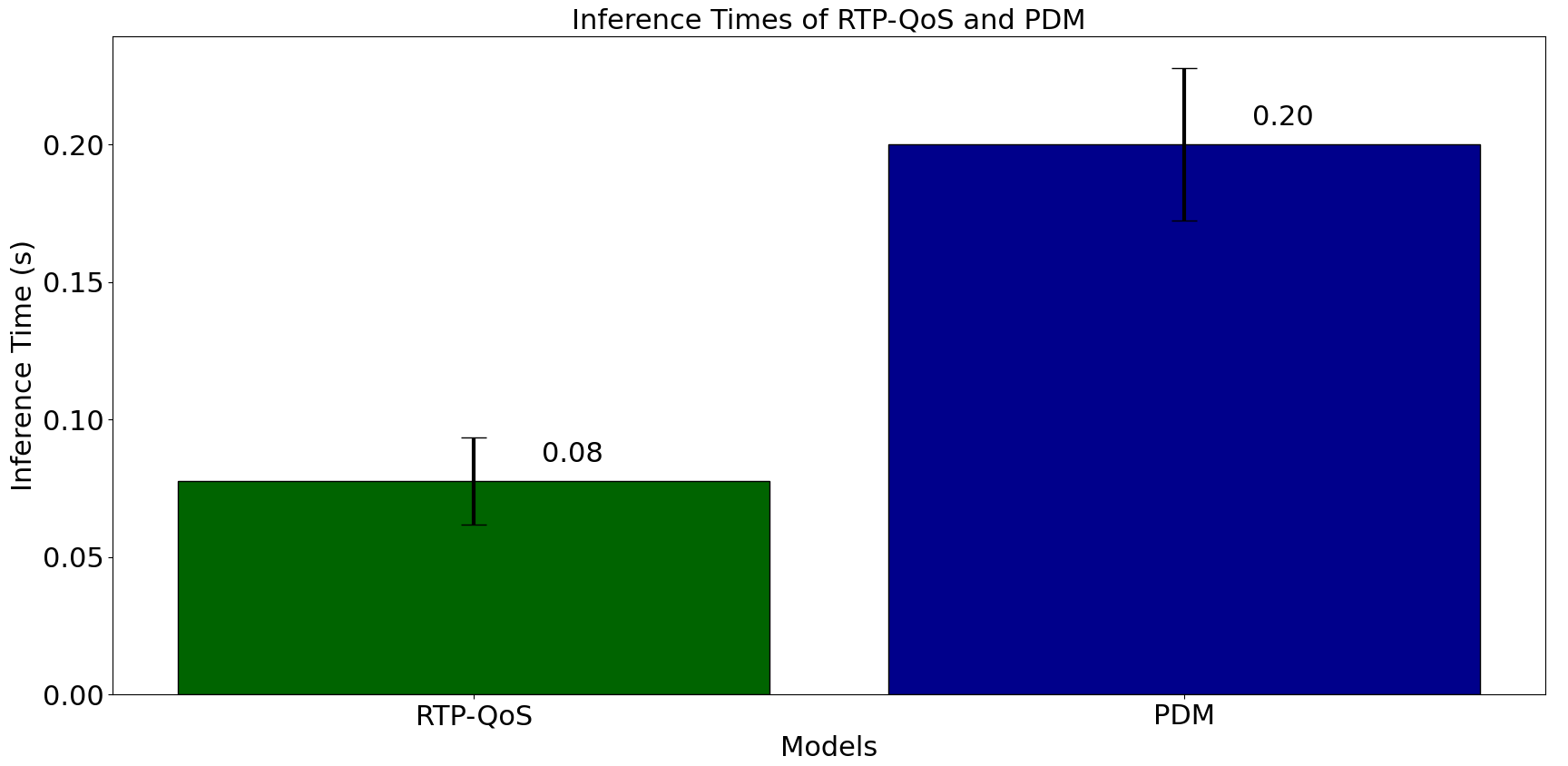}
	\caption{Training (top) and inference (bottom) times of the RTP-QoS and PDM models, which are the components of the proposed IDQS architecture}\label{fig:ComputationTimes}
\end{figure}

In last, we analyze the computation time required for the components, namely the RTP-QoS and the PDM models, of the proposed IDQS architecture. To this end, Figure~\ref{fig:ComputationTimes} presents the results obtained for the training and inference times for these components. 

The measurements in Figure~\ref{fig:ComputationTimes} (top) show that the training of RTP-QoS takes approximately 360 seconds, and that of PDM takes about 87 seconds on average. The inference time measurements in Figure~\ref{fig:ComputationTimes} (bottom) show that the QoS prediction for one sample takes 0.08 seconds on average while a decision about LDoS for a single sample takes 0.2 seconds on average. That is, the total computation time of IDQS for a single decision is $0.28$ seconds on average and its standard deviation is considerably small. {This computational efficiency transforms IDQS from a passive monitor into a proactive defense mechanism. In a real-world IoT gateway deployment, this allows for the activation of mitigation protocols (e.g., dynamic rate limiting) well before the LDoS attack can reach its cumulative exhaustion peak.}


\section{Conclusion} \label{sec:Conclusion}

This paper has proposed a new holistic system for Intrusion Detection via QoS Prediction, in short IDQS. The proposed IDQS is comprised of Recurrent Trend Predictive QoS Predictor (RTP-QoS) and Pairwise Decision Model (PDM), which are operating sequentially. The RTP-QoS model combines the modified version of rTPNN and the QoS prediction layer. PDM consists of a pairwise layer and fully connected layers. The pairwise layer of the PDM learns the relationships between the predicted and actual QoS pairs while the fully connected layers identify the signatures of LDoS attacks by processing the pairwise relationships combined with the predicted QoS. 

The performance of the proposed IDQS and its components were evaluated using the publicly available ``SDN-SlowRate-DDoS'' dataset \cite{dataset}. Evaluation results indicate that IDQS achieves over 0.79 accuracy for majority of text experiments in this dataset through 5-fold cross-validation. Moreover, the performance of the RTP-QoS model was compared to the performances of Long Short-Term Memory (LSTM), Multilayer Perceptron (MLP), Nonlinear Perceptron (NP). The results showed that IDQS consistently outperforms these models, achieving considerably lower prediction error of below 1\% N-SMAPE. The PDM developed was also benchmarked against Feed-Forward Network (FFN). The results indicated that PDM consistently outperforms FFN with more stable and reliable attack detection. Furthermore, IDQS demonstrates computational efficiency, requiring only $0.28$ seconds on average for complete QoS prediction and attack classification, highlighting its suitability for real-time use in resource-limited environments such as IoT systems.

{Despite its high accuracy, the IDQS framework and this paper has some limitations that define its current operational scope. A primary constraint of this work is the scarcity of high-quality, publicly available datasets specifically curated for Low-Rate DoS attacks in IoT environments. While the SDN-SlowRate-DDoS and CIC-IDS2017 datasets provide essential benchmarks, the limited variety of documented low-rate attack vectors restricts the exposure of model to the full spectrum of evolving stealthy threats. In addition, the model relies on a cold-start period determined by the time step $\tau$; for instance, with $\tau=50$ s, the system requires an initial window of historical data before it can generate its first reliable prediction. Furthermore, while the PDM is robust, extreme legitimate network congestion, such as a flash crowd event, could potentially mimic the QoS degradation signature of an LDoS attack, leading to transient false positives. Finally, the current iteration assumes a clean training phase; if the baseline data is unknowingly poisoned with attack traces, the sensitivity of the model to subtle deviations could be diminished.}

Future work shall first focus on evaluating the proposed IDQS in a real-life IoT testbed for analyzing not only the QoS prediction and attack detection performances but also impact of the IDQS on device(s) hosting it. {Such evaluation is crucial also for collecting a high-quality real-life low-rate attack data.} The proposed IDQS shall also be extended to perform multi-class attack detection, aiming to include both slow-rate and intense, such as DoS or DDoS, attacks. In addition, the measurements presented in Section~\ref{sec:CompTimeResults} revealed that RTP-QoS and PDM training took 360~s and 87~s respectively, indicating that IDQS is for further integration with adaptive learning. 

\section*{Acknowledgments}

This work is funded by the project FENG.01.01-IP.01-A0GV/24, ``Advanced Application Maintenance and Diagnostics Tool for IT Startups,''.

Mert Nak{\i}p gratefully acknowledges that his work is supported by the Foundation for Polish Science (FNP) under agreement no. START 057.2025.

\bibliographystyle{elsarticle-num}
\bibliography{ldos_references, qos_lowrate_attacks, predictive_security_ldos, attack_detection, background, bot_detection, datasets, RNN, security_issues_iot, self_supervised, statistics, techniques}

\end{document}